\documentclass[twocolumn,aps,amsmath,amssymb,floatfix,superscriptaddress,prb,longbibliography]{revtex4-1}
\usepackage{graphicx}% Include figure files
\usepackage{dcolumn}% Align table columns on decimal point
\usepackage{bm}% bold math
\usepackage{xspace}
\usepackage{hyperref}
\usepackage{color}
\usepackage[flushleft]{threeparttable}

\def\pmao{PrMgAl$_{11}$O$_{19}$\xspace}

\begin{document}
% \draft command makes pacs numbers print

\title{Possible gapless quantum spin liquid behavior in the triangular-lattice Ising antiferromagnet PrMgAl$_{11}$O$_{19}$}

\author{Zhen~Ma}
\email{zma@hbnu.edu.cn}
\affiliation{Hubei Key Laboratory of Photoelectric Materials and Devices, School of Materials Science and Engineering, Hubei Normal University, Huangshi 435002, China}
\affiliation{State Key Laboratory of Surface Physics and Department of Physics, Fudan University, Shanghai 200433, China}
\author{Shuhan~Zheng}
\author{Yingqi~Chen}
\author{Ruokai~Xu}
\affiliation{Hubei Key Laboratory of Photoelectric Materials and Devices, School of Materials Science and Engineering, Hubei Normal University, Huangshi 435002, China}
\author{Zhao-Yang~Dong}
\email{zhydong@njust.edu.cn}
\affiliation{Department of Applied Physics, Nanjing University of Science and Technology, Nanjing 210094, China}
\author{Jinghui~Wang}
\affiliation{ShanghaiTech Laboratory for Topological Physics and School of Physical Science and Technology, ShanghaiTech University, Shanghai 200031, China}
\author{Hong~Du}
\affiliation{Tsung-Dao Lee Institute $\&$ School of Physics and Astronomy, Shanghai Jiao Tong University, Shanghai 200240, China}
\author{Jan~Peter~Embs}
\affiliation{Laboratory for Neutron Scattering and Imaging, Paul Scherrer Institut, 5232 Villigen, Switzerland}
\author{Shuaiwei~Li}
\author{Yao~Li}
\author{Yongjun~Zhang}
\author{Meifeng~Liu}
\affiliation{Hubei Key Laboratory of Photoelectric Materials and Devices, School of Materials Science and Engineering, Hubei Normal University, Huangshi 435002, China}
\author{Ruidan~Zhong}
\affiliation{Tsung-Dao Lee Institute and School of Physics and Astronomy, Shanghai Jiao Tong University, Shanghai 200240, China}
\author{Jun-Ming~Liu}
\affiliation{National Laboratory of Solid State Microstructures and Department of Physics, Nanjing University, Nanjing 210093, China}
\affiliation{Collaborative Innovation Center of Advanced Microstructures, Nanjing University, Nanjing 210093, China}
\author{Jinsheng~Wen}
\email{jwen@nju.edu.cn}
\affiliation{National Laboratory of Solid State Microstructures and Department of Physics, Nanjing University, Nanjing 210093, China}
\affiliation{Collaborative Innovation Center of Advanced Microstructures, Nanjing University, Nanjing 210093, China}

%\date{\today}

\begin{abstract}
Quantum spin liquids~(QSLs) represent a novel state where spins are highly entangled but do not order even at zero temperature due to strong quantum fluctuations. Such a state is mostly studied in Heisenberg models defined on geometrically frustrated lattices. Here, we turn to a new triangular-lattice antiferromagnet \pmao, in which the interactions are believed to be of Ising type. Magnetic susceptibility measured with an external field along the $c$ axis is two orders of magnitude larger than that with a field in the $ab$ plane, displaying an ideal easy-axis behavior. Meanwhile, there is no magnetic phase transition or spin freezing observed down to 1.8~K. Ultralow-temperature specific heat measured down to 50~mK does not capture any phase transition either, but a hump at 4.5~K, below which the magnetic specific heat exhibits a quasi-quadratic temperature dependence that is consistent with a Dirac QSL state. Inelastic neutron scattering technique is also employed to elucidate the nature of its ground state. In the magnetic excitation spectra, there is a gapless broad continuum at the base temperature 55~mK, in favor of the realization of a gapless QSL. Our results provide a scarce example for the QSL behaviors observed in an Ising-type magnet, which can serve as a promising platform for future research on QSL physics based on an Ising model.
\end{abstract}

\maketitle

\section{Introduction}

Quantum spin liquids~(QSLs) represent a unique quantum disordered state of matter, which can host fractional elementary excitations and emergent gauge structures~\cite{Anderson1973153,nature464_199,RevModPhys.89.025003,Broholmeaay0668}. They have attracted enormous research interest due to the relevance to the understanding of high-temperature superconductivity and promising applications in quantum computation~\cite{anderson1,Kitaev20032,RevModPhys.80.1083,nature464_199,RevModPhys.89.025003,Broholmeaay0668}. To unveil the nature of QSLs, it is of considerable significance to seek and investigate real QSL materials. One route to this goal is to introduce strong quantum fluctuations in a magnetic system~\cite{nature464_199,RevModPhys.89.025003,Broholmeaay0668}, which can be reached via a variety of manners~\cite{doi:10.1146/annurev-matsci-080819-011453,doi:10.1021/acs.chemrev.0c00641}, such as geometrical frustration, spin anisotropy, competition between different paths of magnetic interactions, low dimensionality of magnetic correlations, etc. In this context, quite a few QSL candidate materials have been proposed in recent years~\cite{RevModPhys.89.025003,Broholmeaay0668}.

Among the currently existing QSL candidates, triangular-lattice rare-earth compounds are a representative branch. In rare-earth magnets, the magnetic correlation is generally strong and 4$f$ electrons are quite localized. The spin-orbit-entangled nature of the local moments often brings highly anisotropic magnetic couplings~\cite{RevModPhys.82.53,PhysRevB.94.035107,Liu_2018}. The combination of geometrical frustration and spin anisotropic greatly improves the possibility to realize a QSL state~\cite{PhysRevB.94.035107,Liu_2018,RevModPhys.89.025003}. As such, the triangular-lattice ytterbium-based compound YbMgGaO$_4$ has become a canonical example~\cite{sr5_16419,prl115_167203,nature540_559,np13_117,
PhysRevLett.117.097201,PhysRevLett.118.107202,nc8_15814,
PhysRevLett.122.137201}. In particular, the availability of large single crystals makes it easy to perform various experimental investigations~\cite{prl115_167203}. Earlier measurements including specific heat~\cite{sr5_16419}, inelastic neutron scattering~(INS)~\cite{nature540_559,np13_117}, muon spin relaxation~($\mu$SR)~\cite{nc8_15814}, etc, indeed show positive evidences for a QSL. However, subsequent results of thermal conductivity and ac magnetic susceptibility experiments are at odds with the scenario\cite{PhysRevLett.117.267202,PhysRevLett.120.087201}. The controversial point lies in the structural disorder that nonmagnetic Mg$^{2+}$ and Ga$^{3+}$ are randomly distributed in the same Wyckoff position, which results in complicated effect on the ground state~\cite{PhysRevLett.118.107202,PhysRevLett.120.087201,ZhenMa:106101,PhysRevX.8.031028,npjqm4_12,Broholmeaay0668}. Some people think disorder is detrimental to the formation of QSL states and mimics a QSL~\cite{PhysRevLett.119.157201,PhysRevLett.120.087201,PhysRevX.8.031028,PhysRevB.102.224415,Broholmeaay0668}, while others instead believe it is able to enhance quantum fluctuations~\cite{PhysRevLett.118.107202,https://doi.org/10.1002/qute.201900089,PhysRevB.105.024418,PhysRevLett.115.077001,PhysRevLett.127.017201}. It has been open to question until now. Anyway, seeking new QSL candidates eliminating the influence from disorder is an imperative way. As a result, the disorder-free $A$Yb${Ch}_{\rm 2}$ ($A$ = alkali metal and ${Ch}$ = O, S, Se) compounds are proposed and considered as a perfect investigated regime~\cite{Liu_2018}. Nevertheless, given that alkali metal elements are inherently susceptible to site deficiency as well as challenge in growing high-quality single crystals~\cite{np15_1058,PhysRevMaterials.3.074405,PhysRevX.11.021044,PhysRevB.108.064405}, enlarging the QSL candidate family is highly desirable.

More recently, it reports that rare-earth based hexaaluminates $ReM$Al$_{11}$O$_{19}$, where $Re$ denotes lanthanides and $M$ is Zn or Mg, as a new family of perfect triangular-lattice compounds were synthesized~\cite{ASHTAR2019146,C9TC02643F}. Due to the significant difference in radii of the constituent ions, there is little structural disorder in this series of compounds~\cite{ASHTAR2019146,C9TC02643F}. Moreover, the larger distance of magnetic layers makes it closer to an ideal two-dimensional structure when compared with YbMgGaO$_4$. Among them, PrZnAl$_{11}$O$_{19}$ has firstly caught the attention. It hosts a non-Kramers doublet ground state resulting from strong spin-orbit coupling of 4$f^2$ Pr$^{3+}$ ions and crystal-electric-field effect with $D_{3h}$ symmetry~\cite{ASHTAR2019146,PhysRevB.106.134428}. The $T^2$ behavior of low-temperature specific heat and diffusive excitations of INS point to a gapless QSL, while $\mu$SR experiments also evidence the persistent spin dynamics down to 0.27~K~\cite{PhysRevB.106.134428}. Nonetheless, all the current measurements are confined to its polycrystalline samples, which can only yield powder-average results and thus would miss the crucial information related to crystalline directions, the detailed investigations on its single crystals are still absent at the moment.

Considering the volatile nature upon heating of ZnO in PrZnAl$_{11}$O$_{19}$~\cite{PhysRevLett.120.087201}, it is actually tough to obtain the single crystals. In this work, we turn to its sister compound \pmao, of which the high-quality single crystals have been successfully grown by us. We comprehensively investigate its magnetic properties, specific heat as well as low-energy magnetic excitations via INS at a much lower temperature of several tens of milli-kelvin. It is surprising that the magnetic susceptibility with a field in the $ab$ plane is negligible when compared with that along the $c$ axis, suggesting an ideal Ising model. The fitting to the inverse magnetic susceptibility data gives rise to a Curie-Weiss~($\Theta_{\rm CW}$) temperature of -8.1~K, reflecting the dominant antiferromagnetic interactions in this system. Moreover, the absence of bifurcation between zero-field-cooling~(ZFC) and field-cooling~(FC) data excludes the possibility of spin freezing. Specific heat measurements down to as low as 50~mK reveal no sharp $\lambda$-shaped peak indicative of a phase transition, but a hump around 4.5~K, below which the data fit well with a quasi-quadratic temperature dependence. It is reminiscent of a gapless Dirac QSL. INS experiments capture broad spin excitation continuum below 2.5~meV, evidencing the possible realization of a gapless QSL state. Our work provides a promising platform to study QSL physics in a real Ising system.

\section{Experimental Details}

Polycrystalline samples of \pmao were synthesized by solid-state reaction method. The raw materials Pr$_2$O$_3$~(99.9\%), MgO~(99.99\%), and Al$_2$O$_3$~(99.99\%) were weighed with a stoichiometric ratio. Then the precursor powders were mixed and thoroughly ground in an agate mortar. Finally, they were loaded into an alumina crucible and sintered at 1550~$^\circ$C for 60 hours with an intermediate grinding. High-quality single crystals were grown in air via a two-mirror optical floating-zone furnace~(MF-2400, Cyberstar Corp.). During the growth, the seed and feed rods spun in an opposite direction with a speed of 25 r/min, and meanwhile traveled downwards with a speed of 9 mm/h. The obtained single crystals are chartreuse with a size up to 7$\times$3$\times$1.0~mm$^3$ for a piece, as shown in the inset of Fig.~\ref{fig1}(c).

Powder x-ray diffraction data were collected at room temperature in an x-ray diffractometer (X$^\prime$TRA, ARL) using the Cu-$K_\alpha$ edge with a wavelength of 1.54~\AA. In the measurements, the scan range of 2$\theta$ is from 10$^\circ$ to 80$^\circ$ with a step of 0.02$^\circ$ and a rate of 10$^\circ$/min, respectively. Rietveld refinements on powder XRD data were performed by GSAS software. X-ray backscattering Laue~(Photonic Science) pattern was taken when the beam was perpendicular to the cleavage plane~($ab$ plane) of the single crystal and the exposure time is 60~s. Dc magnetic susceptibility was measured on a piece of single crystal with the typical mass $\sim$15.9~mg. This procedure was performed within 1.8-300~K in a Quantum Design physical property measurement system~(PPMS, Dynacool), equipped with a vibrating sample magnetometer option. Specific heat was measured on a 4.3-mg single crystal in a PPMS Dynacool equipped with a dilution refrigerator, so that the lowest temperature can reach 50~mK. INS experiments were carried out on FOCUS, a time-flight spectrometer located at Paul Scherrer Institute~(PSI) at Villigen, Switzerland. About 5-g sample was loaded into a pure copper can in a dilution refrigerator and it was able to cool down to a base temperature of 55~mK. The incident neutron wavelength was fixed as $\lambda \sim$ 4.7~\AA, corresponding an good energy resolution of $\Delta E$ = 0.05~meV (half width at half maximum, HWHM). For each temperature, we collect data for about 12 h.

\section{Results}
\subsection{Crystal structure and x-ray diffraction}

\begin{figure}[htb]
\centerline{\includegraphics[width=0.98\linewidth]{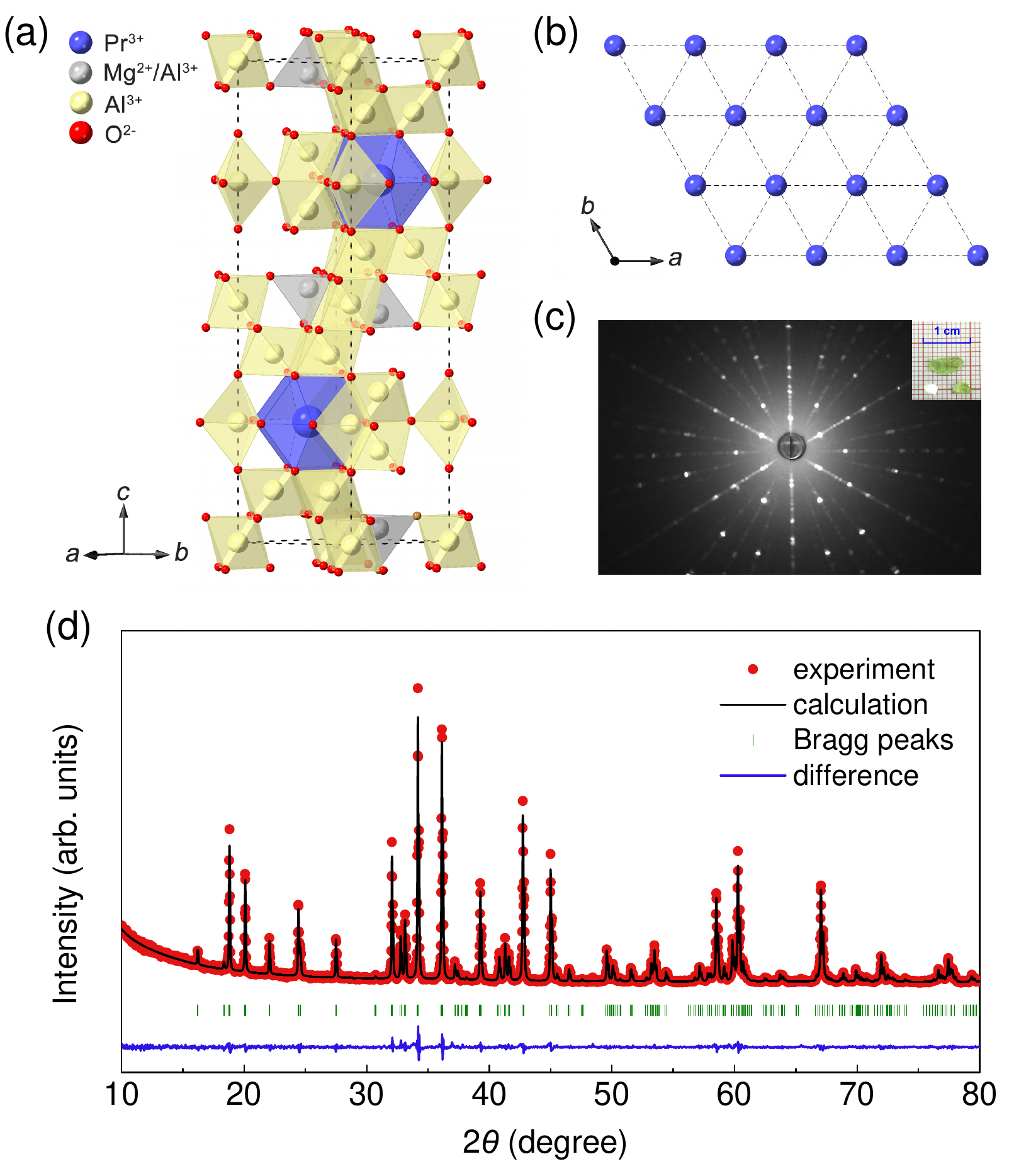}}
\caption{
(a) Schematic crystal structure of \pmao. (b) Top view of the triangular layer consisting of magnetic Pr$^{3+}$ ions. Dashed lines in (a) and (b) denote the crystalline unit cell and the triangular unit, respectively. (c) X-ray Laue pattern of the (001) plane of a single crystal. Inset shows a photograph of chartreuse single crystals with several typical pieces. (d) Rietveld refinement results for the powder XRD data collected at room temperature. Red circles and black solid line indicate the measured data and calculated results upon Rietveld refinement, respectively. Green ticks denote Bragg peak positions of \pmao with space group P6$_3$/mmc. Blue solid line represents the difference between the experiments and calculations for this compound.
\label{fig1}}
\end{figure}

Figures~\ref{fig1}(a) and \ref{fig1}(b) show the schematics of the crystal structure and a two-dimensional triangular layer of \pmao, respectively. Magnetic Pr$^{3+}$ ions form edge-shared triangular layers that are well separated by several nonmagnetic layers consisting of AlO$_6$ octahedra, AlO$_5$ hexahedra, and Al/MgO$_4$ tetrahedra, featuring an ideal two-dimensional nature of magnetic couplings~\cite{ASHTAR2019146,C9TC02643F}. Along the $c$ axis, the magnetic triangular layers have an $ABAB$ stacking arrangement. It is emphasized that although the nonmagnetic Al$^{3+}$ ions located at the center of tetrahedra suffer from 1/2 occupation mixed with Mg$^{2+}$ ions, we believe it should have little impact on the magnetic ground state, since the antisite disorder only exists inside Al/MgO$_4$ tetrahedra and the ratio of this configuration of polyhedra is relatively less in the structural unit cells. Moreover, the nonmagnetic layers with this disorder problem are not directly adjacent to the magnetic layers, either. We will discuss it more in the discussion part later.

In Fig.~\ref{fig1}(d), we present the powder XRD patterns as well as the Rietveld refinement results of this compound. It can be seen that the polycrystalline sample is a single-phase material, with no additional impurity peaks observed in the whole measured range. This is vital for the further preparation of its single crystals. The refinements were performed based on a hexagonal structure in the space group P6$_3$/mmc~\cite{ASHTAR2019146,C9TC02643F}. It gives rise to the lattice parameters $a$ = $b$ = 5.5846(3)~\AA, $c$ = 21.8730(12)~\AA, and $\alpha$ = $\beta =$ 90$^\circ$, $\gamma$ = 120$^\circ$. From this result, we can further calculate that the separations of the nearest intralayer and interlayer Pr$^{3+}$ ions are 5.5846~\AA~and 11.9582~\AA, respectively. It means the strength of magnetic interplay within the magnetic triangular layers is actually $\sim$10 times larger than that outside the triangular layers when considering the dipolar energy goes as 1/$r^3$~(Ref.~\onlinecite{PhysRevLett.84.3430}). It justifies the two-dimensional nature of this compound as told above. These parameters are also in good agreement with the previous report~\cite{ASHTAR2019146}. The final refinement parameters are $R_{\rm p} \approx$ 8.97\%, $R_{\rm wp} \approx$ 7.14\%, and $\chi^2 \approx$ 1.485, respectively, suggesting a high purity of our synthesized samples. Upon these well-prepared polycrystalline samples, we have succeeded in growing the single crystals of \pmao. X-ray backscattering Laue pattern was taken to characterize them when the incident beam was perpendicular to the cleavage plane, as shown in Fig.~\ref{fig1}(c). We can clearly see the sharp diffraction spots with a six-fold symmetry, reflecting the structural characteristic of a triangular framework. The diffraction partten also shows some sporadic intensity near the strong spots. It indicates that there is a small domain of the single crystal with its crystallographic orientation in the $ab$ plane somewhat deviating from the main one. It should have ignorable impact on the following measured results, since such a phenomenon is very weak and almost all the experimental investigations were performed along the crystallographic $c$ axis or upon powder sample.

\subsection{Magnetic susceptibility}

\begin{figure}[htb]
\centerline{\includegraphics[width=0.98\linewidth]{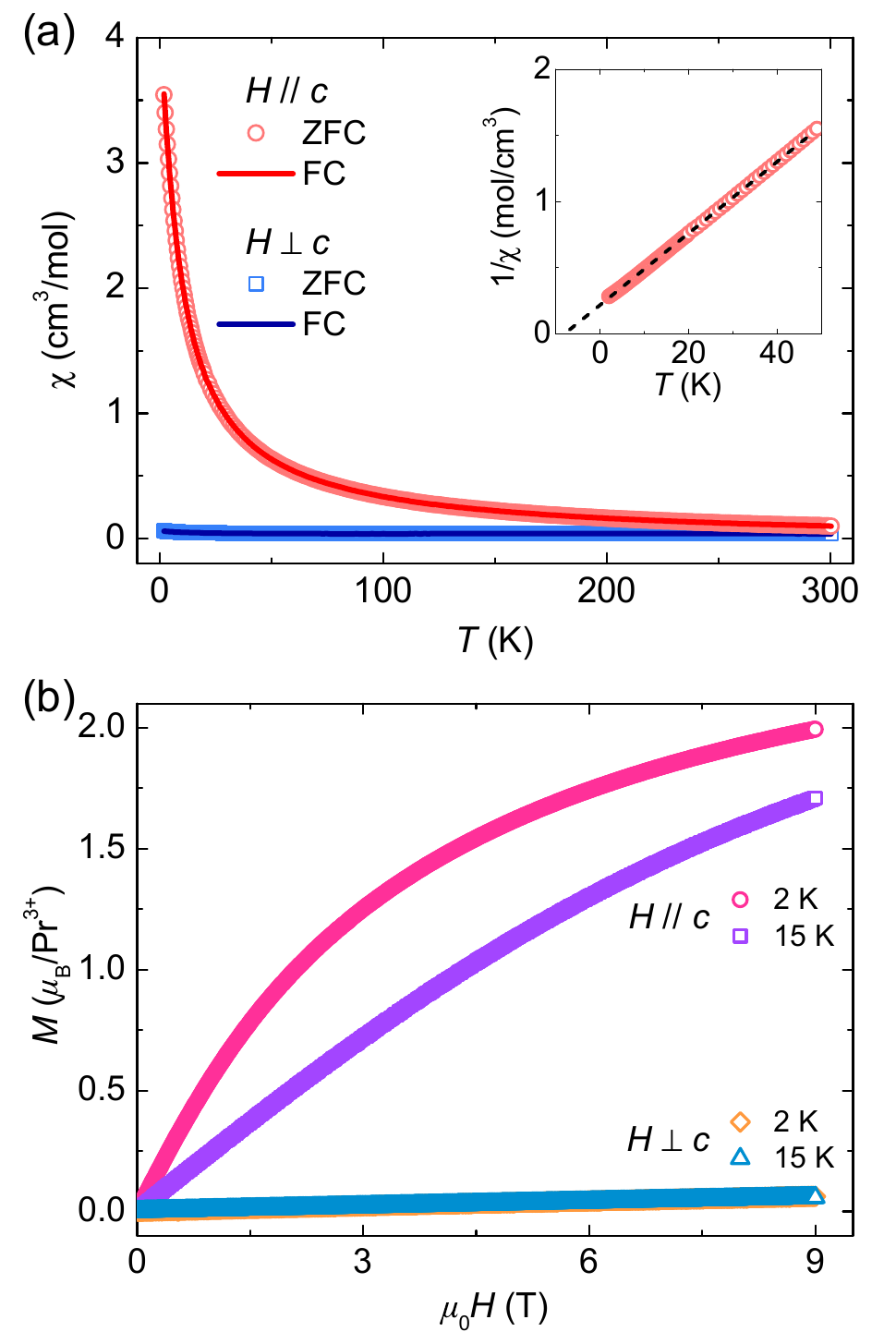}}
\caption{
 (a) Temperature dependence of the magnetic susceptibility~($\chi$) for \pmao single crystals with the external fields parallel and perpendicular to the $c$ axis, respectively. The measurements were performed in both zero-field-cooling~(ZFC) and field-cooling~(FC) conditions with a field of 0.1~T down to 1.8~K. The inset shows the inverse magnetic susceptibility data with the field along $c$ axis. The dashed line is the fit with the Curie-Weiss law. (b) Magnetic-field dependence of the magnetization for the single-crystal sample with two different directions at 2~K and 15~K, respectively.
\label{fig2}}
\end{figure}

We then characterize the magnetic properties of \pmao single crystals by measuring dc magnetic susceptibility~($\chi$) at a small field of 0.1~T, and the results are shown in Fig.~\ref{fig2}(a). The susceptibility with an external field along the $c$ axis is two orders of magnitude larger than that with a field applied in the $ab$ plane. It signals that \pmao is an experimental realization of a triangular-lattice Ising model with the $c$ axis as its easy axis of magnetization. More intriguing, there is no any magnetic transition or anomaly captured in the whole measurement process, which indicates the paramagnetic behavior persisting to the lowest temperature of 1.8~K. We speculate that the low-temperature paramagnetic state should be modulated by quantum fluctuations, since the thermal fluctuations are proportional to temperature and it should have little impact on the ground state when down to as low as 1.8~K. Besides that, no signature of splitting between ZFC and FC data can be distinguished in both directions, ruling out the possibility of spin freezing down to the lowest temperature available. Such a behavior is also observed in its polycrystalline sample as well as its sister compound PrZnAl$_{11}$O$_{19}$~\cite{ASHTAR2019146,PhysRevB.106.134428}. In general, the Curie-Weiss fitting to low-temperature data can reveal the intrinsic physics related to its ground state, while the electrons will be thermally populated to excited crystal-electric-field~(CEF) levels at higher temperatures, and thus the inverse magnetic susceptibility deviates from the temperature-linear behavior due to CEF effect~\cite{ASHTAR2019146}. In the inset of Fig.~\ref{fig2}(a), we present the inverse low-temperature susceptibility data along the $c$ axis that can be well fitted by the Curie-Weiss law. It yields a Curie-Weiss temperature $\Theta_{\rm CW}\sim$~-8.1~K within a typical fitting range of 1.8-50~K, indicating the dominant antiferromagnetic nature of the exchange interactions between the Pr$^{3+}$ local moments in this system. This value is close to that of -6.4~K determined upon powder data~\cite{ASHTAR2019146}. It is also noteworthy that the intralayer Pr$^{3+}$-Pr$^{3+}$ distance is $\sim$5.584~\AA~from the above structural characterization, which is larger than those of the magnetic ions in other referenced compounds. For instance, it is $\sim$3.401~\AA~for Yb$^{3+}$-Yb$^{3+}$ separation in the heavily-studied YbMgGaO$_4$, of which the Curie-Weiss temperature is about -4~K for the powder sample~\cite{sr5_16419}. However, the main path of the magnetic interactions in \pmao is a nearly straight ``Pr$^{3+}$-O$^{2-}$-Pr$^{3+}$" bridge~\cite{ASHTAR2019146}, whose length is $\sim$5.590~\AA~determined by us, while it is a polygonal ``Yb$^{3+}$-O$^{2-}$-Yb$^{3+}$" one with the length of $\sim$4.456~\AA~in the latter~\cite{nature540_559,sr5_16419}. That means the effective distance of the superexchange paths is actually comparable for these two compounds. On the other hand, the relatively large effective moment of Pr$^{3+}$ ($\sim$2.96~$\mu_{\rm {eff}}$~\cite{ASHTAR2019146}) compared with Yb$^{3+}$ ($\sim$~2.6$\mu_{\rm {eff}}$~\cite{PhysRevB.99.180401}) at low temperatures would further strengthen the magnetic interactions in this regime. Thus, it is reasonable to yield such a value of Curie-Weiss temperature even though there is relatively large distance between the intralayer magnetic ions in \pmao. For the magnetic susceptibility data perpendicular the $c$ axis, it cannot be fitted by the Curie-weiss law, since the spins in this compound exhibit an ideal easy-axis behavior and there is almost no component of spin moments along the hard-axis direction, as shown in Fig.~\ref{fig2}(a).

The isothermal field-dependent magnetization up to 9~T with two different directions is depicted in Fig.~\ref{fig2}(b). The magnetization with the field perpendicular to the $c$ axis is insignificant compared with that along the $c$ axis, which is in line with the Ising-type magnetic behavior determined in magnetic susceptibility. For the case of field along the $c$ axis at 2~K, the magnetization curve progressively increases in a small-field range. Then it turns to deviate from the linear response regime around 1.5~T, which corresponds to the relatively low energy scale of the spin interactions in the ground state. Subsequently, it enters into a smooth transition regime and hardly shows the sign of saturation even though the external field increases up to 9~T, where the maximum magnetization value $\sim$2.0~$\mu_{\rm B}$/Pr$^{3+}$ is far away from 3.2~$\mu_{\rm B}$/Pr$^{3+}$ expected for free Pr$^{3+}$ ions~\cite{ASHTAR2019146}. It means that higher external field is needed to fully polarize the magnetic moments. With increasing the temperature to 15~K, the magnetic couplings between the Pr$^{3+}$ ions are gradually suppressed and the excited crystal field states would be thermally activated, which further keeps the magnetization from saturation and thus makes the magnetization curve remains mostly linear within the current field range.

\subsection{Specific heat}

\begin{figure}[h!]
  \centering
  \includegraphics[width=0.93\linewidth]{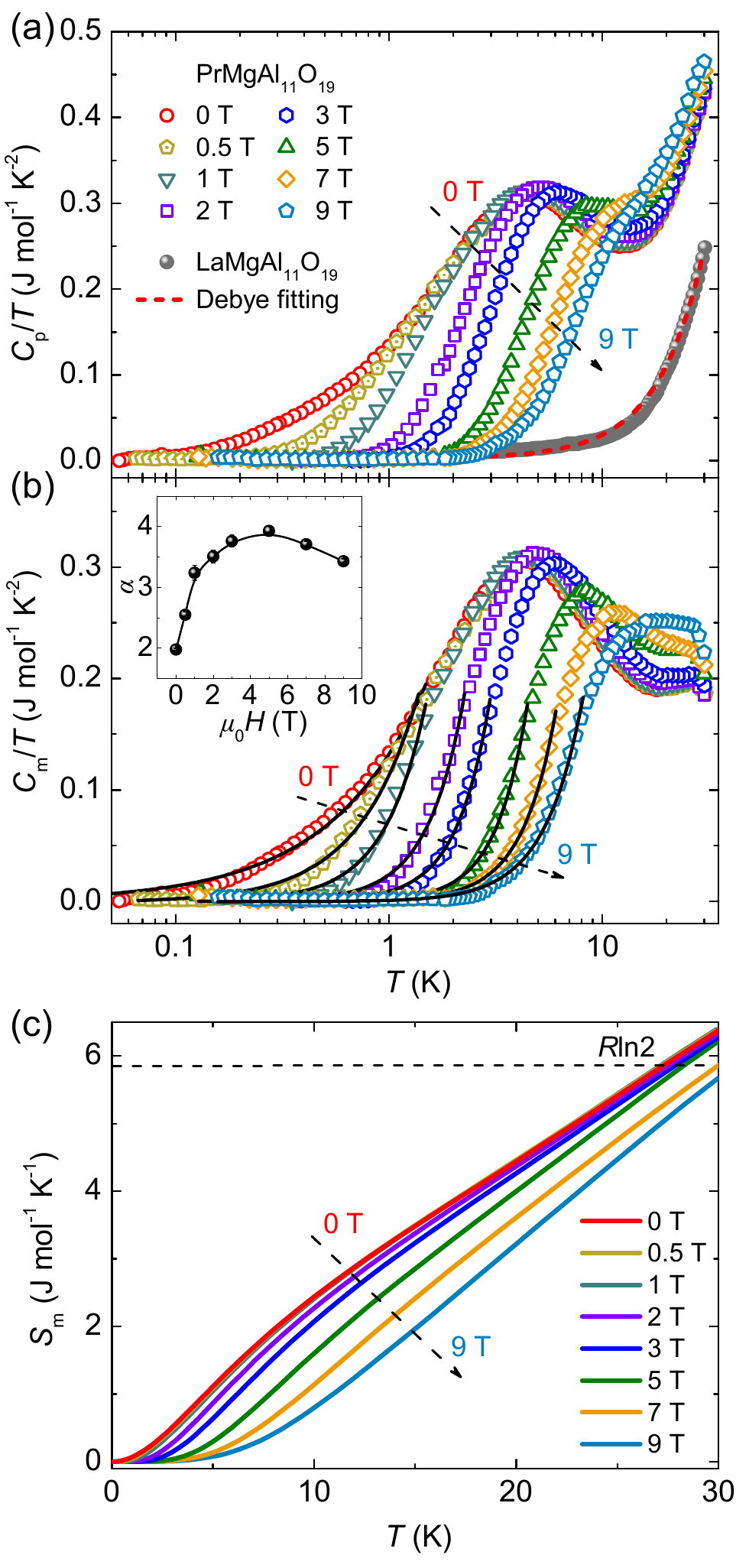}
  \caption{
  (a) Ultralow-temperature specific heat~($C_{\rm p}$) results of \pmao at various magnetic fields. The specific heat of a nonmagnetic reference compound LaMgAl$_{11}$O$_{19}$ at zero field is also shown for comparison, which can be nicely fitted by a Debye model as $C_{\rm p}\propto T^3$. The dashed line denotes such a fitting. (b) Magnetic specific heat~($C_{\rm m}$) of \pmao obtained by subtracting the lattice contribution using an isostructural nonmagnetic sample LaMgAl$_{11}$O$_{19}$. The solid lines denote the fit to the data with a power-law function $C_{\rm m}/T\propto T^{\alpha-1}$. The inset shows the extracted exponent values of $\alpha$ at different fields. The solid in the inset is the guide to the eye. (c) Temperature dependence of magnetic entropy~($S_{\rm m}$) obtained by integrating $C_{\rm m}/T$ data.}
  \label{fig3}
\end{figure}

Above magnetic susceptibility and magnetization characterizations on the single crystals provide the basic information of the magnetic moments and the magnetic interactions in \pmao. To reveal the properties of low-energy magnetic excitations, we perform the specific heat measurements down to an ultralow temperature of 50~mK. In Fig.~\ref{fig3}(a), we depict the specific heat results of \pmao as well as a nonmagnetic reference compound LaMgAl$_{11}$O$_{19}$ for comparison. At zero field, there is no sharp $\lambda$-shaped peak indicative of a phase transition, but a hump around 4.5~K. Such a feature is commonly observed in quite a few QSL candidates, such as triangular-lattice materials $\kappa$-(BEDT-TTF)$_2$Cu$_2$(CN)$_3$~\cite{np4_459} and YbMgGaO$_4$~\cite{sr5_16419}, and the kagome-lattice compound ZnCu$_3$(OH)$_6$Cl$_2$~\cite{prl98_107204}, etc. It may suggest an establishment of short-range correlations while the Schottky anomaly in a two-level system is also proposed~\cite{np4_459,sr5_16419,prl98_107204,PhysRevB.102.224415}. Considering there has been no consensus on its exact physical origin so far, we cannot make a solid conclusion at the moment. However, no matter which case it is, the long-range magnetic transition can be excluded. When an external field is applied on this regime, the hump is gradually suppressed and it shifts towards higher temperatures, reflecting the continuous modulation of magnetic ground state via the external fields. To reveal the possible novel excitations in the ground state, it is helpful to perform the quantitative analysis on its magnetic specific heat~($C_{\rm m}$) data. In general, the total specific heat of a crystallized system with gapless elementary excitations can be represented as $C_{\rm p} = \gamma T + \beta T^3$~(Refs.~\onlinecite{np4_459,nc2_275}), where the first term can originate from itinerant electrons, spinons or other quasiparticles with fractional spin while the second one may result from phonon and/or magnon excitations. As a result, the magnetic specific heat of \pmao can be obtained by subtracting the total specific heat of an isostructural reference compound LaMgAl$_{11}$O$_{19}$ [also shown in the Fig.~\ref{fig3}(a)], which can be taken as the pure lattice part of the target compound~\cite{C9TC02643F}.

Figure~\ref{fig3}(b) shows the results of magnetic specific heat of \pmao at various fields. There is indeed no phase transition observed, confirming the absence of long-range magnetic order down to at least 50~mK, two orders of magnitude lower than its interaction energy scale of $|\Theta_{\rm CW}| \sim$ 8.1~K. More importantly, the low-temperature data can be well fitted by a power-law function of $C_{\rm m}/T = AT^{\alpha-1}$, implying the promising nontrivial excitations in the ground state. At zero field, it gives rise to a $\alpha$ value of 1.98(2) within the fitting range between 0.05 and 1.5~K. The result of $\alpha$ approaching to 2 is unusual for an insulator and it may evidence the gapless Dirac QSL state with Dirac nodes~\cite{PhysRevLett.98.117205,PhysRevLett.123.207203}. In fact, such a behavior has already been observed in a wider range of QSL candidates, such as the diluted honeycomb-lattice Kitaev regime $\alpha$-Ru$_{0.8}$Ir$_{0.2}$Cl$_3$~\cite{PhysRevLett.124.047204}, kagome-lattice antiferromagnet YCu$_3$(OH)$_6$Br$_2$[Br$_x$(OH)$_{1-x}$]~\cite{PhysRevB.105.024418,PhysRevB.105.L121109}, square-kagome compound Na$_6$Cu$_7$BiO$_4$(PO$_4)_4$~\cite{PhysRevB.105.155153}, as well as the triangular-lattice compound YbZn$_2$GaO$_5$~\cite{arXiv_2305.20040}. Nevertheless, it is the first time to capture this feature in an Ising-type QSL candidate. When a magnetic field is applied along the $c$ axis, the magnetic specific heat is quickly suppressed, indicating the effective modulation of the magnetic ground state by the external field. We have also extracted the exponent of $\alpha$ at other fields in this way, and the result is shown in the inset of Fig.~\ref{fig3}(b). It can be seen that $\alpha$ increases dramatically with the field below around 3~T, followed by a saturation value close to 4 at the moderate fields. As the field continues increasing up to 9~T, it falls back to 3 to some extent. This nonmonotonic response of $\alpha$ to the increasing field is also observed in its sister compound PrZnAl$_{11}$O$_{19}$ and it was attributed to the inhomogeneous effective magnetic field throughout the polycrystalline sample when considering different orientations of crystallographic axes for each grain inside the polycrystal~\cite{PhysRevB.106.134428}. However, current results are obtained based on high-quality single crystals. Thus, the reason of the polycrystalline sample itself can be excluded. It should reflect more complex intrinsic physics. Considering the significant decrease of magnetic specific heat with the applied magnetic field, the occurrence of field-induced spin gap that annihilates partial low-energy magnetic excitations may be responsible for the behavior. It deserves more detailed investigations on the magnetic-field effect in the future. Figure~\ref{fig3}(c) shows the temperature dependence of magnetic entropy by integrating the $C_{\rm m}/T$ data presented in Fig.~\ref{fig3}(b). At zero field, the released magnetic entropy from the lowest measured temperature of 50 mK to 30 K is about $R$ln2, where $R$ is the ideal gas constant. It shows no sign of saturation within the current temperature range, just like the case of its sister compound PrZnAl$_{11}$O$_{19}$ (Ref.~\onlinecite{PhysRevB.106.134428}). The calculated residual magnetic entropy at 50 mK is less than 0.0001$\%$$R$ln2. The near-zero value of residual magnetic entropy rules out the possibility of magnetic transition at lower temperatures. Thus, our low-temperature result can reveal the magnetic properties related to the ground state. It is consistent with the feature of a QSL. When applying a magnetic field into this regime, the magnetic entropy is continuously reduced, indicating the suppression of spin fluctuations by the external field.

\subsection{INS spectra}

\begin{figure}[htb]
  \centering
  \includegraphics[width=0.98\linewidth]{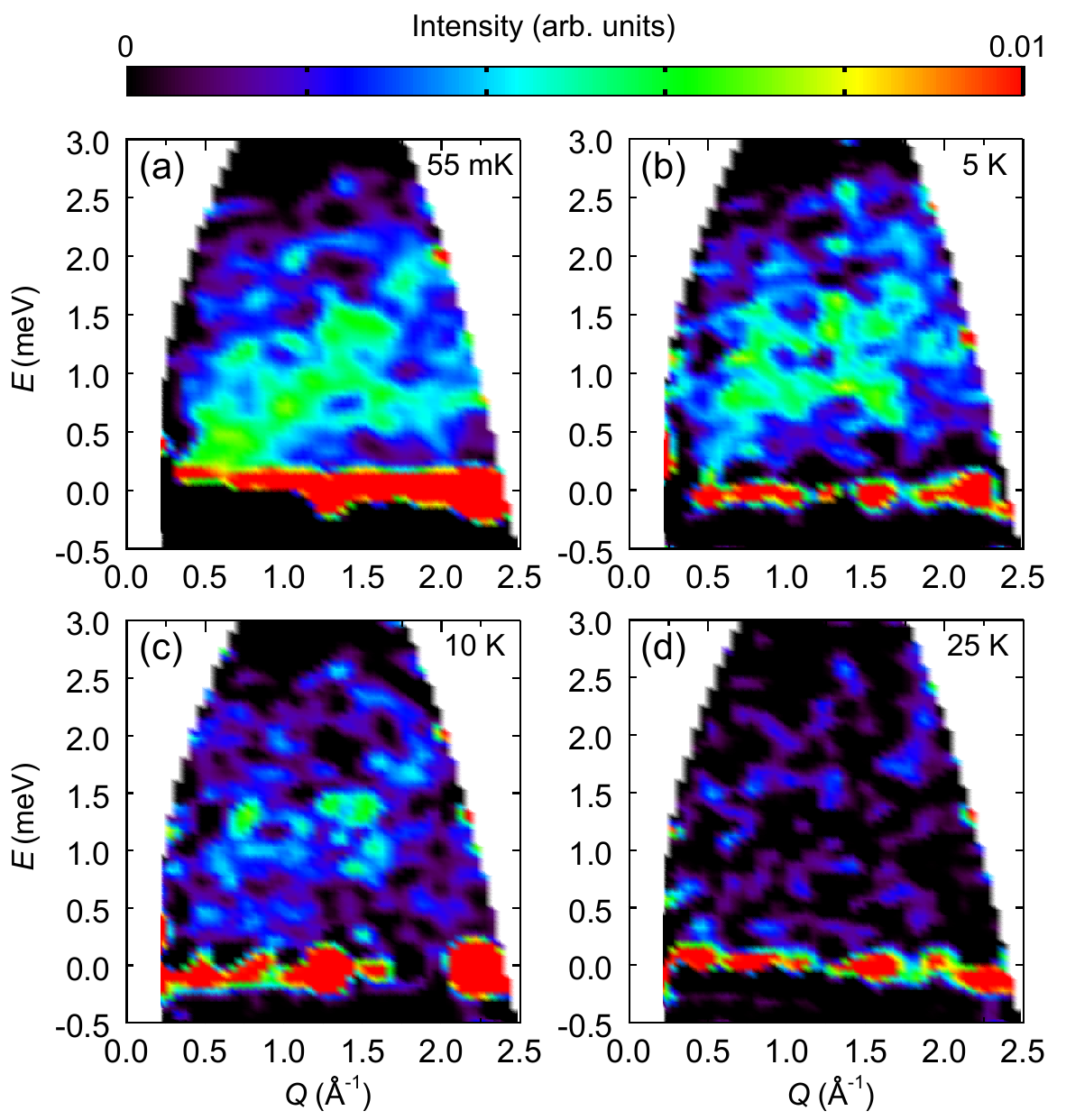}
  \caption{
  (a)-(d) Inelastic neutron scattering spectra of \pmao measured with an incident neutron energy $E_{\rm i}$~=~3.7~meV at several temperatures. To distinguish the possible low-energy magnetic excitations covered by the strong signals from low-lying crystal electric field excitations, the data at a higher temperature of 50~K were used for background subtraction.}
  \label{fig4}
\end{figure}

\begin{figure*}[htb]
  \centering
  \includegraphics[width=6.8in]{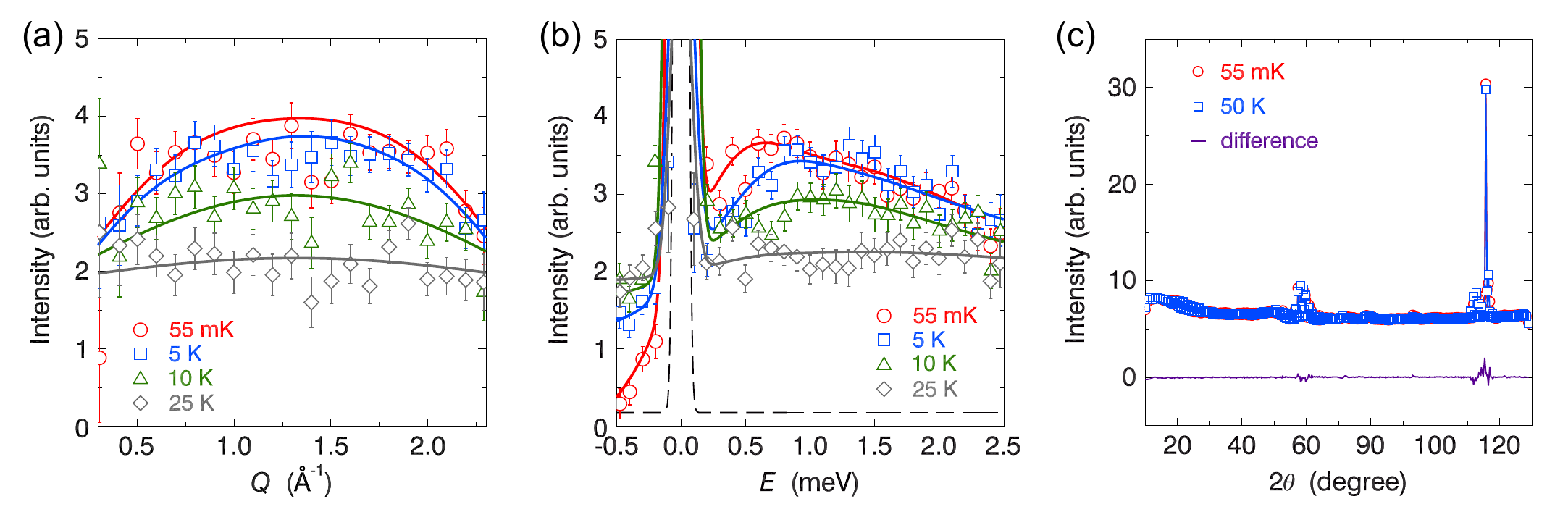}
  \caption{
  (a) Wave-vector $\bm{Q}$ dependence of the energy-integrated neutron scattering intensity in the range from 0.25 to 2.5~meV at several temperatures. (b) Energy dependence of the integrated intensity in the $\bm{Q}$ range from 0 to 2.5~\AA$^{-1}$. The solid lines in (a) and (b) are guides to the eye. The dashed line in (b) denotes the instrumental resolution. (c) Elastic neutron scattering data were obtained by integrating the intensity in an energy window of [-0.1, 0.1] meV. Errors of the neutron scattering data represent one standard deviation.}
  \label{fig5}
\end{figure*}

To further elucidate the nature of low-energy magnetic excitations, INS experiments were performed at several characteristic temperatures. In fact, we have observed a low-lying CEF level around 1.5 meV (not shown), which is the same as that measured in its sister compound PrZnAl$_{11}$O$_{19}$~\cite{PhysRevB.106.134428}. However, given that this low-lying CEF level may mask the possible magnetic excitation signals, just like the cases in some other rare-earth QSL candidates with non-Kramers ions, such as TbInO$_3$ and Tm$_3$Sb$_3$Mg$_2$O$_{14}$~\cite{np15_262,PhysRevB.102.224415}, the high-temperature data of 50~K were used for background subtraction. Although at 50~K the CEF level of 1.5~meV would be occupied by the thermally-populated electrons from the ground state, what we concern most is if there are any underlying low-energy magnetic excitations below the strong CEF excitations. One route to this goal is to subtract a high-temperature constant data to pick it out. Thus, it should be fine whether taking the data of 50~K or a lower temperature, such as 10~K as the background, since they both represent an invariant constant that needs to be subtracted.  The so-obtained results are shown in Fig.~\ref{fig4}. Aside from the strong elastic scattering signals around zero energy associated with crystal structure, there are diffusive excitations captured below 2.5~meV, which are in sharp contrast to the well-defined spin-wave excitations expected in a magnetically ordered regime. Therefore, INS results rule out the possibility of long-range magnetic order at an ultralow temperature of 55~mK. The broad continuum of spin excitations has been also observed in a few frustrated quantum magnets and it is considered as a hallmark of QSLs~\cite{RevModPhys.89.025003,Broholmeaay0668}. It should be true for \pmao when taking into account the above result of quasi-quadratic temperature dependence of magnetic specific heat indicative of a Dirac QSL state. These diffusive excitations weaken with the increasing temperatures and nearly disappear at 25~K. The evident temperature dependence of spectral intensities confirms its magnetic nature. We also need to point out that the intenisity in the elastic region is different at different temperatures, just as shown in Fig.~\ref{fig4}. It results from the small intensity scale displayed in the figure. In fact, there is a huge variation of intensity between the measured elastic and inelastic signals. That means the statistical error of the elastic signal may be even larger than the inelastic intensity itself. Here, what we care about most is the inelastic scattering region and the intensity scale is thus defined to a small window. It would significantly affect the image of the elastic region.

To investigate the low-energy excitations in detail, we integrate the magnetic spectral weight with energy ranging from 0.25 to 2.5~meV and plot the intensities as a function of wave-vector $\bm{Q}$, as shown in Fig.~\ref{fig5}(a). It is clear that there are higher scattering intensities centered around the intermediate $\bm{Q}$ range for the low-temperature data, while the intensities at both small and large momentum-transfer positions are relatively low. This feature is different from the CEF excitations where the excitation intensity has a monotonic $\bm{Q}$ dependence~\cite{PhysRevLett.118.107202,PhysRevB.98.134401,PhysRevB.106.134428}, proving the origin of low-energy magnetic excitations. When the temperature increases to 25~K, the broad peak becomes indistinguishable, suggesting the complete exhaustion of magnetic excitations due to the predominant role of thermal fluctuations at this temperature. We also integrate the spectral intensities all over $\bm{Q}$ and obtain the energy $E$ dependence of intensities at various temperatures, as shown in Fig.~\ref{fig5}(b). The intensities on the neutron-energy-loss ($E>0$) side have the same temperature dependence as that of $\bm{Q}$, affirming the magnetic nature of spectral weight. Moreover, the low-energy magnetic excitations are very close to and difficult to be distinguished from the strong elastic line. In other words, \pmao may be a gapless spin system, which is consistent with what we deduce from specific heat measurements. To further clarify whether there is a gap or not, INS experiment with higher energy resolution is required. The magnetic excitations gradually weaken and finally vanish above approximately 2.5~meV. In contrast, the temperature dependence of intensities on the energy-gain ($E<0$) side are opposite due to detailed balance~\cite{prl98_107204}. Similar phenomena are also observed in other QSL candidates, such as Zn$_3$Cu(OH)$_6$Cl$_2$~\cite{prl98_107204}, Ce$_2$Zr$_2$O$_7$~\cite{PhysRevLett.122.187201}, and NaRuO$_2$~\cite{np19_949}. Such a behavior of contrasting temperature dependence of excitation intensities emphasizes the reliability of our experimental data. It should be noted that the integrated intensity in Figs.~\ref{fig5}(a) and \ref{fig5}(b) is affected by the selected area of the INS spectra shown in Fig.~\ref{fig4}, but the overall behaviors of $\bm{Q}$ and energy dependence of intensity would not be changed. Based on the original measured magnetic spectra (no background subtraction), the integrated intensities stemmed from the elastic channels are as well extracted and shown in Fig.~\ref{fig5}(c). The neutron diffraction data of 55~mK and 50~K are overlapped and there is no additional magnetic Bragg peak observed, especially in the low-angle range where magnetic peaks generally show up for a magnetically ordered system. This result agrees well with the absence of magnetic order determined in both magnetic susceptibility and specific heat, pointing to a fascinating QSL ground state in this Ising magnet.

\subsection{CEF analysis}

To further facilitate our understanding of its ground state, especially to clarify the issue of singlet or doublet in a low-symmetric CEF environment of the non-Kramers ion Pr$^{3+}$, we also perform a CEF analysis based on the method of point-charge approximation. Pr$^{3+}$ has an electron configuration of 4$f^2$, in which the spin and orbital angular momentums ($S$ = 1, $L$ = 5) are entangled into a total angular momentum $J$ = 4 due to the strong spin-orbit coupling. The corresponding 9-fold degenerate states are further split under the CEF effect. The effective CEF Hamiltonian is obtained by the point-charge model according to the $D_{3h}$ point-group symmetry~\cite{HUTCHINGS1964227},
\begin{equation}\label{CEF}
  H_{\rm CEF}=\sum_{n,m}B_n^m O_n^m,
\end{equation}
where $O_n^m$ and $B_n^m$ are the Steven operators~\cite{Stevens_1952} and CEF parameters, respectively. It turns out that there are two lowest singlets with an extremely small gap of 0.00011~meV, which would compose a nearly degenerate non-Kramers doublet ground state, as shown in Table~\ref{CEFl}. These two singlet levels are $|+\rangle=-0.3458|-3\rangle+0.6168|-1\rangle-0.6168|1\rangle+0.3458|3\rangle$ and $|-\rangle=0.0177|-4\rangle-0.4098|-2\rangle+0.8145|0\rangle-0.4098|2\rangle+0.0177|4\rangle$, respectively.

\begin{table*}[htb]
  \begin{threeparttable}
\caption{Calculated CEF energy levels obtained by the point-charge approximation.}
\label{CEFl}
\begin{tabular*}{\textwidth}{@{\extracolsep{\fill}}cccccccccc}
\hline \hline
\begin{minipage}{2cm}\vspace{1mm} Eigenvalues \vspace{1mm} \end{minipage} & $E_1$ & $E_2$ & $E_3$ & $E_4$ & $E_5$ & $E_6$ & $E_7$ & $E_8$ & $E_9$  \\
\hline
 Value~(meV) & 0.00000 & 0.00011 & 29.31636 & 30.63402 & 30.63419 & 40.66725 & 40.66754 & 41.25604 & 49.15044  \\
\hline \hline
\end{tabular*}
\end{threeparttable}
\end{table*}

Unlike Kramers doublets, for this non-Kramers doublet, the longitudinal spin component behaves as the magnetic dipole moment, while the transverse spin component behaves as the magnetic quadrupole moment. This anisotropy of the magnetic moments implies a possible Ising-type magnetic behavior. While the current CEF results merely depending on symmetry analysis cannot match up the excited CEF levels, such as the energy level of 1.5~meV captured by INS, it is able to qualitatively identify the low-lying ground state, since the non-Kramers doublet is mainly determined by the point-group symmetry~\cite{Stevens_1952,EDVARDSSON1998230,PhysRevB.98.045119}. Thus, it is sufficient to simply do the symmetry analysis despite the lack of experimental CEF data to optimize the parameters for the effective CEF Hamiltonian. It is also worth noting that such a tiny gap of 0.00011 meV between the quasidoublet ground state is only a rough value rather than a quite accurate one due to the accuracy lamination of the point-charge calculations~\cite{PhysRevResearch.3.023012}. Further fitting upon full CEF levels measured by both low and high-energy INS experiments would give rise to more accurate CEF parameters. Nevertheless, the components of the non-Kramers doublet would not change significantly. In fact, the case that non-Kramers ions in a low-symmetric CEF environment create a quasidoublet ground state is commonly observed in other non-Kramers compounds, such as PrTiNbO$_6$~\cite{PhysRevB.97.184434}, Pr$_3$BWO$_9$~\cite{nagl2024excitation}, Tb$_3$Sb$_3$Mg$_2$O$_{14}$ and Ho$_3$Sb$_3$Mg$_2$O$_{14}$~\cite{PhysRevResearch.3.023012}, etc. It is thus reasonable to identify the ground state as a quasidoublet state in \pmao.

\section{Discussions}

Currently, the relationship between structural disorder and the QSL state has become a hotly debated issue~\cite{RevModPhys.89.025003,Broholmeaay0668,npjqm4_12,ZhenMa:106101,PhysRevLett.127.267202}. The prototypical examples include a kagome-lattice material Zn$_3$Cu(OH)$_6$Cl$_2$ and a triangular-lattice compound YbMgGaO$_4$. Strong disorder effect hinders the direct insight into the intrinsic ground states of these candidate materials. However, unlike the case in Zn$_3$Cu(OH)$_6$Cl$_2$ where antisite disorder directly occurs between magnetic Cu$^{2+}$ and nonmagnetic Zn$^{2+}$ ions~\cite{RevModPhys.88.041002}, there is no such mixed occupation in \pmao, but only weak disorder between the nonmagnetic Al$^{3+}$ and Mg$^{2+}$ ions~\cite{ASHTAR2019146}. Although this problem is quite analogous to the case of YbMgGaO$_4$ where nonmagnetic Ga$^{3+}$ and Mg$^{2+}$ ions are randomly distributed at the same Wyckoff position~\cite{sr5_16419,np13_117}, we should stress that the disorder content in this material is much less than that in YbMgGaO$_4$, since there is no structural disorder in AlO$_6$ and AlO$_5$ polyhedra and site mixing only exists in Al/MgO$_4$ tetrahedra which are in minority in the structure of \pmao~\cite{ASHTAR2019146,PhysRevB.106.134428}. On the other hand, the nonmagnetic layers with structural disorder in YbMgGaO$_4$ are directly adjacent to the magnetic Yb$^{3+}$ layers, which has a significant influence on the ground state~\cite{PhysRevLett.119.157201,PhysRevLett.120.087201,PhysRevX.8.031028,PhysRevB.102.224415}. In contrast, there is an intermediate nonmagnetic layer consisting of disorder-free AlO$_6$ octahedra between the magnetic Pr$^{3+}$ layer and the antisite-disorder Al/MgO$_4$ layer that can effectively block their correlations. Thus, we argue that weak disorder effect in \pmao should have ignorable impact on the ground state and all our measured results can reflect the intrinsic physics. Additionally, while the observed CEF level of 1.5~meV is close to the quasidoublet ground state, as presented by the above CEF analysis, we believe the low-temperature behavior of quasi-quadratic temperature dependence of magnetic specific heat can effectively reveal the magnetic properties associated with its ground state, since it is measured below 1~K ($\sim$0.09 meV) that is far away from the contaminations from this CEF level. To quantitatively study the influence of the CEF effect on milli-kelvin specific heat, one can introduce nonmagnetic ions of La$^{3+}$ into \pmao to dilute its magnetism, so that the single-ion magnetism related to the CEF splitting would be separated out when the content of La$^{3+}$ reaches a very high level, since the ingredient of cooperative magnetism resulting from the inter-site magnetic couplings is excluded at this time, like the cases of Tm$_{1-x}$Lu$_x$MgGaO$_4$~\cite{PhysRevX.10.011007} and Pr$_{1-x}$La$_x$TiNbO$_6$~\cite{PhysRevB.97.184434}. The low-lying CEF levels will also affect the low-energy magnetic excitations related to the ground state, which is commonly observed in other compounds with non-Kramers ions, such as TbInO$_3$~\cite{np15_262} and Tm$_3$Sb$_3$Mg$_2$O$_{14}$~\cite{PhysRevB.102.224415}. It is thus highly desirable to perform a systematical CEF analysis based on full excited levels measured by low and high-energy INS in the future. In this case, the information of CEF splitting can be accurately learnt, which will help thoroughly unveil the intrinsic magnetic properties of the ground state. Given these, we believe the present results have well fulfilled our goal to propose a scarce example of an Ising magnet that exhibits QSL behaviors in a perfect triangular-lattice compound \pmao.

In fact, there is a relevant research on its sister compound PrZnAl$_{11}$O$_{19}$, which shows similar behaviors to \pmao~\cite{PhysRevB.106.134428}. Given that the experimental results are quite similar for these two isostrutural compounds, we speculate that they should share the same QSL ground state. In other words, PrZnAl$_{11}$O$_{19}$ may be another Ising system that hosts a QSL state. In the meanwhile, we also notice that there have already been a few other works suggesting possible QSL states in an Ising model, such as the triangular-lattice compounds NdTa$_7$O$_{19}$ and ErMgGaO$_4$~\cite{nm21_416,PhysRevB.101.094432}. However, the single-crystal sample of the former is still lacking for now while the latter suffers from the same disorder problem as YbMgGaO$_4$~\cite{CEVALLOS20185}. Compared with them, \pmao may serve as an ideal platform to explore the QSL physics in an Ising magnet. Especially, the availability of large-size and high-quality single crystals of this material will make it more favorable. Besides the experimental investigations, there is a more recent numerical calculation work that also supports the possible realization of a QSL state in a triangular-lattice Ising regime~\cite{arXiv_2307.03545}. The growing research interests in theory and experiment both advance the QSL investigations from a conventional Heisenberg model to a promising Ising one.

\section{Summary}

In summary, we have succeeded in growing single crystals for a perfect triangular-lattice magnet \pmao. The magnetic susceptibility characterizations reveal an ideal Ising system with the $c$ axis as its easy axis of magnetization. Meanwhile, there is no long-range magnetic order nor spin freezing observed down to 1.8~K. Specific heat measurements confirm the absence of magnetic order at a much too low temperature of 50~mK. More importantly, the behavior of a quasi-quadratic temperature dependence of magnetic specific points to a Dirac QSL state, which can be further modulated by the external field. Moreover, the broad continuum of low-energy magnetic excitation spectra is captured by INS experiments, which is also consistent with a gapless QSL state. Our work provides a rare example of a real Ising system that exhibits QSL behaviors, which serves as a fertile ground to explore the nontrivial quantum states in an Ising system.

\section{Acknowledgments}

We thank Tong Chen for helpful discussions. The work was supported by the National Key Projects for Research and Development of China with Grant No.~2021YFA1400400, National Natural Science Foundation of China with Grants No.~12204160, No.~12225407, No.~12074174, No.~12074111, and No.~52272108, Hubei Provincial Natural Science Foundation of China with Grant No.~2023AFA105, and Fundamental Research Funds for the Central Universities. We thank the neutron beam time from PSI with Proposal No.~20202284 and the support from the Sample Environment Group for setting up the dilution refrigerator on FOCUS.

%\bibliography{topo}

\begin{thebibliography}{67}%
\makeatletter
\providecommand \@ifxundefined [1]{%
 \@ifx{#1\undefined}
}%
\providecommand \@ifnum [1]{%
 \ifnum #1\expandafter \@firstoftwo
 \else \expandafter \@secondoftwo
 \fi
}%
\providecommand \@ifx [1]{%
 \ifx #1\expandafter \@firstoftwo
 \else \expandafter \@secondoftwo
 \fi
}%
\providecommand \natexlab [1]{#1}%
\providecommand \enquote  [1]{``#1''}%
\providecommand \bibnamefont  [1]{#1}%
\providecommand \bibfnamefont [1]{#1}%
\providecommand \citenamefont [1]{#1}%
\providecommand \href@noop [0]{\@secondoftwo}%
\providecommand \href [0]{\begingroup \@sanitize@url \@href}%
\providecommand \@href[1]{\@@startlink{#1}\@@href}%
\providecommand \@@href[1]{\endgroup#1\@@endlink}%
\providecommand \@sanitize@url [0]{\catcode `\\12\catcode `\$12\catcode
  `\&12\catcode `\#12\catcode `\^12\catcode `\_12\catcode `\%12\relax}%
\providecommand \@@startlink[1]{}%
\providecommand \@@endlink[0]{}%
\providecommand \url  [0]{\begingroup\@sanitize@url \@url }%
\providecommand \@url [1]{\endgroup\@href {#1}{\urlprefix }}%
\providecommand \urlprefix  [0]{URL }%
\providecommand \Eprint [0]{\href }%
\providecommand \doibase [0]{http://dx.doi.org/}%
\providecommand \selectlanguage [0]{\@gobble}%
\providecommand \bibinfo  [0]{\@secondoftwo}%
\providecommand \bibfield  [0]{\@secondoftwo}%
\providecommand \translation [1]{[#1]}%
\providecommand \BibitemOpen [0]{}%
\providecommand \bibitemStop [0]{}%
\providecommand \bibitemNoStop [0]{.\EOS\space}%
\providecommand \EOS [0]{\spacefactor3000\relax}%
\providecommand \BibitemShut  [1]{\csname bibitem#1\endcsname}%
\let\auto@bib@innerbib\@empty
%</preamble>
\bibitem [{\citenamefont {Anderson}(1973)}]{Anderson1973153}%
  \BibitemOpen
  \bibfield  {author} {\bibinfo {author} {\bibfnamefont {P.W.}\ \bibnamefont
  {Anderson}},\ }\bibfield  {title} {\enquote {\bibinfo {title} {{Resonating
  valence bonds: A new kind of insulator?}}}\ }\href@noop {} {\bibfield
  {journal} {\bibinfo  {journal} {Mater. Res. Bull.}\ }\textbf {\bibinfo
  {volume} {8}},\ \bibinfo {pages} {153} (\bibinfo {year}
  {1973})}\BibitemShut {NoStop}%
\bibitem [{\citenamefont {Balents}(2010)}]{nature464_199}%
  \BibitemOpen
  \bibfield  {author} {\bibinfo {author} {\bibfnamefont {Leon}\ \bibnamefont
  {Balents}},\ }\bibfield  {title} {\enquote {\bibinfo {title} {{Spin liquids
  in frustrated magnets}},}\ }\href {\doibase 10.1038/nature08917} {\bibfield
  {journal} {\bibinfo  {journal} {Nature}\ }\textbf {\bibinfo {volume} {464}},\
  \bibinfo {pages} {199} (\bibinfo {year} {2010})}\BibitemShut {NoStop}%
\bibitem [{\citenamefont {Zhou}\ \emph {et~al.}(2017)\citenamefont {Zhou},
  \citenamefont {Kanoda},\ and\ \citenamefont {Ng}}]{RevModPhys.89.025003}%
  \BibitemOpen
  \bibfield  {author} {\bibinfo {author} {\bibfnamefont {Yi}~\bibnamefont
  {Zhou}}, \bibinfo {author} {\bibfnamefont {Kazushi}\ \bibnamefont {Kanoda}},
  \ and\ \bibinfo {author} {\bibfnamefont {Tai-Kai}\ \bibnamefont {Ng}},\
  }\bibfield  {title} {\enquote {\bibinfo {title} {{Quantum spin liquid
  states}},}\ }\href {\doibase 10.1103/RevModPhys.89.025003} {\bibfield
  {journal} {\bibinfo  {journal} {Rev. Mod. Phys.}\ }\textbf {\bibinfo {volume}
  {89}},\ \bibinfo {pages} {025003} (\bibinfo {year} {2017})}\BibitemShut
  {NoStop}%
\bibitem [{\citenamefont {Broholm}\ \emph {et~al.}(2020)\citenamefont
  {Broholm}, \citenamefont {Cava}, \citenamefont {Kivelson}, \citenamefont
  {Nocera}, \citenamefont {Norman},\ and\ \citenamefont
  {Senthil}}]{Broholmeaay0668}%
  \BibitemOpen
  \bibfield  {author} {\bibinfo {author} {\bibfnamefont {C.}~\bibnamefont
  {Broholm}}, \bibinfo {author} {\bibfnamefont {R.~J.}\ \bibnamefont {Cava}},
  \bibinfo {author} {\bibfnamefont {S.~A.}\ \bibnamefont {Kivelson}}, \bibinfo
  {author} {\bibfnamefont {D.~G.}\ \bibnamefont {Nocera}}, \bibinfo {author}
  {\bibfnamefont {M.~R.}\ \bibnamefont {Norman}}, \ and\ \bibinfo {author}
  {\bibfnamefont {T.}~\bibnamefont {Senthil}},\ }\bibfield  {title} {\enquote
  {\bibinfo {title} {Quantum spin liquids},}\ }\href {\doibase
  10.1126/science.aay0668} {\bibfield  {journal} {\bibinfo  {journal}
  {Science}\ }\textbf {\bibinfo {volume} {367}},\ \bibinfo {pages} {eaay0668}
  (\bibinfo {year} {2020})}\BibitemShut {NoStop}%
\bibitem [{\citenamefont {Anderson}(1987)}]{anderson1}%
  \BibitemOpen
  \bibfield  {author} {\bibinfo {author} {\bibfnamefont {P.~W.}\ \bibnamefont
  {Anderson}},\ }\bibfield  {title} {\enquote {\bibinfo {title} {{The
  Resonating Valence Bond State in La$_2$CuO$_4$ and Superconductivity}},}\
  }\href@noop {} {\bibfield  {journal} {\bibinfo  {journal} {Science}\ }\textbf
  {\bibinfo {volume} {235}},\ \bibinfo {pages} {1196} (\bibinfo {year}
  {1987})}\BibitemShut {NoStop}%
\bibitem [{\citenamefont {Kitaev}(2003)}]{Kitaev20032}%
  \BibitemOpen
  \bibfield  {author} {\bibinfo {author} {\bibfnamefont {A.~Yu.}\ \bibnamefont
  {Kitaev}},\ }\bibfield  {title} {\enquote {\bibinfo {title} {Fault-tolerant
  quantum computation by anyons},}\ }\href {\doibase
  http://dx.doi.org/10.1016/S0003-4916(02)00018-0} {\bibfield  {journal}
  {\bibinfo  {journal} {Ann. Phys.}\ }\textbf {\bibinfo {volume} {303}},\
  \bibinfo {pages} {2} (\bibinfo {year} {2003})}\BibitemShut {NoStop}%
\bibitem [{\citenamefont {Nayak}\ \emph {et~al.}(2008)\citenamefont {Nayak},
  \citenamefont {Simon}, \citenamefont {Stern}, \citenamefont {Freedman},\ and\
  \citenamefont {Das~Sarma}}]{RevModPhys.80.1083}%
  \BibitemOpen
  \bibfield  {author} {\bibinfo {author} {\bibfnamefont {Chetan}\ \bibnamefont
  {Nayak}}, \bibinfo {author} {\bibfnamefont {Steven~H.}\ \bibnamefont
  {Simon}}, \bibinfo {author} {\bibfnamefont {Ady}\ \bibnamefont {Stern}},
  \bibinfo {author} {\bibfnamefont {Michael}\ \bibnamefont {Freedman}}, \ and\
  \bibinfo {author} {\bibfnamefont {Sankar}\ \bibnamefont {Das~Sarma}},\
  }\bibfield  {title} {\enquote {\bibinfo {title} {{Non-Abelian anyons and
  topological quantum computation}},}\ }\href {\doibase
  10.1103/RevModPhys.80.1083} {\bibfield  {journal} {\bibinfo  {journal} {Rev.
  Mod. Phys.}\ }\textbf {\bibinfo {volume} {80}},\ \bibinfo {pages}
  {1083} (\bibinfo {year} {2008})}\BibitemShut {NoStop}%
\bibitem [{\citenamefont {Clark}\ and\ \citenamefont
  {Abdeldaim}(2021)}]{doi:10.1146/annurev-matsci-080819-011453}%
  \BibitemOpen
  \bibfield  {author} {\bibinfo {author} {\bibfnamefont {Lucy}\ \bibnamefont
  {Clark}}\ and\ \bibinfo {author} {\bibfnamefont {Aly~H.}\ \bibnamefont
  {Abdeldaim}},\ }\bibfield  {title} {\enquote {\bibinfo {title} {{Quantum Spin
  Liquids from a Materials Perspective}},}\ }\href {\doibase
  10.1146/annurev-matsci-080819-011453} {\bibfield  {journal} {\bibinfo
  {journal} {Annu. Rev. Mater. Res.}\ }\textbf {\bibinfo {volume}
  {51}},\ \bibinfo {pages} {495} (\bibinfo {year} {2021})}\BibitemShut {NoStop}%
\bibitem [{\citenamefont {Chamorro}\ \emph {et~al.}(2021)\citenamefont
  {Chamorro}, \citenamefont {McQueen},\ and\ \citenamefont
  {Tran}}]{doi:10.1021/acs.chemrev.0c00641}%
  \BibitemOpen
  \bibfield  {author} {\bibinfo {author} {\bibfnamefont {Juan~R.}\ \bibnamefont
  {Chamorro}}, \bibinfo {author} {\bibfnamefont {Tyrel~M.}\ \bibnamefont
  {McQueen}}, \ and\ \bibinfo {author} {\bibfnamefont {Thao~T.}\ \bibnamefont
  {Tran}},\ }\bibfield  {title} {\enquote {\bibinfo {title} {{Chemistry of
  Quantum Spin Liquids}},}\ }\href {\doibase 10.1021/acs.chemrev.0c00641}
  {\bibfield  {journal} {\bibinfo  {journal} {Chem. Rev.}\ }\textbf
  {\bibinfo {volume} {121}},\ \bibinfo {pages} {2898.} (\bibinfo {year}
  {2021})}\BibitemShut {NoStop}%
\bibitem [{\citenamefont {Gardner}\ \emph {et~al.}(2010)\citenamefont
  {Gardner}, \citenamefont {Gingras},\ and\ \citenamefont
  {Greedan}}]{RevModPhys.82.53}%
  \BibitemOpen
  \bibfield  {author} {\bibinfo {author} {\bibfnamefont {Jason~S.}\
  \bibnamefont {Gardner}}, \bibinfo {author} {\bibfnamefont {Michel J.~P.}\
  \bibnamefont {Gingras}}, \ and\ \bibinfo {author} {\bibfnamefont {John~E.}\
  \bibnamefont {Greedan}},\ }\bibfield  {title} {\enquote {\bibinfo {title}
  {Magnetic pyrochlore oxides},}\ }\href {\doibase 10.1103/RevModPhys.82.53}
  {\bibfield  {journal} {\bibinfo  {journal} {Rev. Mod. Phys.}\ }\textbf
  {\bibinfo {volume} {82}},\ \bibinfo {pages} {53} (\bibinfo {year}
  {2010})}\BibitemShut {NoStop}%
\bibitem [{\citenamefont {Li}\ \emph {et~al.}(2016{\natexlab{a}})\citenamefont
  {Li}, \citenamefont {Wang},\ and\ \citenamefont {Chen}}]{PhysRevB.94.035107}%
  \BibitemOpen
  \bibfield  {author} {\bibinfo {author} {\bibfnamefont {Yao-Dong}\
  \bibnamefont {Li}}, \bibinfo {author} {\bibfnamefont {Xiaoqun}\ \bibnamefont
  {Wang}}, \ and\ \bibinfo {author} {\bibfnamefont {Gang}\ \bibnamefont
  {Chen}},\ }\bibfield  {title} {\enquote {\bibinfo {title} {{Anisotropic spin
  model of strong spin-orbit-coupled triangular antiferromagnets}},}\ }\href
  {\doibase 10.1103/PhysRevB.94.035107} {\bibfield  {journal} {\bibinfo
  {journal} {Phys. Rev. B}\ }\textbf {\bibinfo {volume} {94}},\ \bibinfo
  {pages} {035107} (\bibinfo {year} {2016}{\natexlab{a}})}\BibitemShut
  {NoStop}%
\bibitem [{\citenamefont {Liu}\ \emph {et~al.}(2018{\natexlab{a}})\citenamefont
  {Liu}, \citenamefont {Zhang}, \citenamefont {Ji}, \citenamefont {Liu},
  \citenamefont {Li}, \citenamefont {Wang}, \citenamefont {Lei}, \citenamefont
  {Chen},\ and\ \citenamefont {Zhang}}]{Liu_2018}%
  \BibitemOpen
  \bibfield  {author} {\bibinfo {author} {\bibfnamefont {Weiwei}\ \bibnamefont
  {Liu}}, \bibinfo {author} {\bibfnamefont {Zheng}\ \bibnamefont {Zhang}},
  \bibinfo {author} {\bibfnamefont {Jianting}\ \bibnamefont {Ji}}, \bibinfo
  {author} {\bibfnamefont {Yixuan}\ \bibnamefont {Liu}}, \bibinfo {author}
  {\bibfnamefont {Jianshu}\ \bibnamefont {Li}}, \bibinfo {author}
  {\bibfnamefont {Xiaoqun}\ \bibnamefont {Wang}}, \bibinfo {author}
  {\bibfnamefont {Hechang}\ \bibnamefont {Lei}}, \bibinfo {author}
  {\bibfnamefont {Gang}\ \bibnamefont {Chen}}, \ and\ \bibinfo {author}
  {\bibfnamefont {Qingming}\ \bibnamefont {Zhang}},\ }\bibfield  {title}
  {\enquote {\bibinfo {title} {{Rare-Earth Chalcogenides: A Large Family of
  Triangular Lattice Spin Liquid Candidates}},}\ }\href {\doibase
  10.1088/0256-307x/35/11/117501} {\bibfield  {journal} {\bibinfo  {journal}
  {Chin. Phys. Lett.}\ }\textbf {\bibinfo {volume} {35}},\ \bibinfo {pages}
  {117501} (\bibinfo {year} {2018}{\natexlab{a}})}\BibitemShut {NoStop}%
\bibitem [{\citenamefont {Li}\ \emph {et~al.}(2015{\natexlab{a}})\citenamefont
  {Li}, \citenamefont {Liao}, \citenamefont {Zhang}, \citenamefont {Li},
  \citenamefont {Jin}, \citenamefont {Ling}, \citenamefont {Zhang},
  \citenamefont {Zou}, \citenamefont {Pi}, \citenamefont {Yang}, \citenamefont
  {Wang}, \citenamefont {Wu},\ and\ \citenamefont {Zhang}}]{sr5_16419}%
  \BibitemOpen
  \bibfield  {author} {\bibinfo {author} {\bibfnamefont {Yuesheng}\
  \bibnamefont {Li}}, \bibinfo {author} {\bibfnamefont {Haijun}\ \bibnamefont
  {Liao}}, \bibinfo {author} {\bibfnamefont {Zhen}\ \bibnamefont {Zhang}},
  \bibinfo {author} {\bibfnamefont {Shiyan}\ \bibnamefont {Li}}, \bibinfo
  {author} {\bibfnamefont {Feng}\ \bibnamefont {Jin}}, \bibinfo {author}
  {\bibfnamefont {Langsheng}\ \bibnamefont {Ling}}, \bibinfo {author}
  {\bibfnamefont {Lei}\ \bibnamefont {Zhang}}, \bibinfo {author} {\bibfnamefont
  {Youming}\ \bibnamefont {Zou}}, \bibinfo {author} {\bibfnamefont
  {Li}~\bibnamefont {Pi}}, \bibinfo {author} {\bibfnamefont {Zhaorong}\
  \bibnamefont {Yang}}, \bibinfo {author} {\bibfnamefont {Junfeng}\
  \bibnamefont {Wang}}, \bibinfo {author} {\bibfnamefont {Zhonghua}\
  \bibnamefont {Wu}}, \ and\ \bibinfo {author} {\bibfnamefont {Qingming}\
  \bibnamefont {Zhang}},\ }\bibfield  {title} {\enquote {\bibinfo {title}
  {{Gapless quantum spin liquid ground state in the two-dimensional spin-1/2
  triangular antiferromagnet YbMgGaO$_4$}},}\ }\href@noop {} {\bibfield
  {journal} {\bibinfo  {journal} {Sci. Rep.}\ }\textbf {\bibinfo {volume}
  {5}},\ \bibinfo {pages} {16419} (\bibinfo {year}
  {2015}{\natexlab{a}})}\BibitemShut {NoStop}%
\bibitem [{\citenamefont {Li}\ \emph {et~al.}(2015{\natexlab{b}})\citenamefont
  {Li}, \citenamefont {Chen}, \citenamefont {Tong}, \citenamefont {Pi},
  \citenamefont {Liu}, \citenamefont {Yang}, \citenamefont {Wang},\ and\
  \citenamefont {Zhang}}]{prl115_167203}%
  \BibitemOpen
  \bibfield  {author} {\bibinfo {author} {\bibfnamefont {Yuesheng}\
  \bibnamefont {Li}}, \bibinfo {author} {\bibfnamefont {Gang}\ \bibnamefont
  {Chen}}, \bibinfo {author} {\bibfnamefont {Wei}\ \bibnamefont {Tong}},
  \bibinfo {author} {\bibfnamefont {Li}~\bibnamefont {Pi}}, \bibinfo {author}
  {\bibfnamefont {Juanjuan}\ \bibnamefont {Liu}}, \bibinfo {author}
  {\bibfnamefont {Zhaorong}\ \bibnamefont {Yang}}, \bibinfo {author}
  {\bibfnamefont {Xiaoqun}\ \bibnamefont {Wang}}, \ and\ \bibinfo {author}
  {\bibfnamefont {Qingming}\ \bibnamefont {Zhang}},\ }\bibfield  {title}
  {\enquote {\bibinfo {title} {{Rare-Earth Triangular Lattice Spin Liquid: A
  Single-Crystal Study of ${\mathrm{YbMgGaO}}_{4}$}},}\ }\href@noop {}
  {\bibfield  {journal} {\bibinfo  {journal} {Phys. Rev. Lett.}\ }\textbf
  {\bibinfo {volume} {115}},\ \bibinfo {pages} {167203} (\bibinfo {year}
  {2015}{\natexlab{b}})}\BibitemShut {NoStop}%
\bibitem [{\citenamefont {Shen}\ \emph {et~al.}(2016)\citenamefont {Shen},
  \citenamefont {Li}, \citenamefont {Wo}, \citenamefont {Li}, \citenamefont
  {Shen}, \citenamefont {Pan}, \citenamefont {Wang}, \citenamefont {Walker},
  \citenamefont {Steffens}, \citenamefont {Boehm}, \citenamefont {Hao},
  \citenamefont {Quintero-Castro}, \citenamefont {Harriger}, \citenamefont
  {Frontzek}, \citenamefont {Hao}, \citenamefont {Meng}, \citenamefont {Zhang},
  \citenamefont {Chen},\ and\ \citenamefont {Zhao}}]{nature540_559}%
  \BibitemOpen
  \bibfield  {author} {\bibinfo {author} {\bibfnamefont {Yao}\ \bibnamefont
  {Shen}}, \bibinfo {author} {\bibfnamefont {Yao-Dong}\ \bibnamefont {Li}},
  \bibinfo {author} {\bibfnamefont {Hongliang}\ \bibnamefont {Wo}}, \bibinfo
  {author} {\bibfnamefont {Yuesheng}\ \bibnamefont {Li}}, \bibinfo {author}
  {\bibfnamefont {Shoudong}\ \bibnamefont {Shen}}, \bibinfo {author}
  {\bibfnamefont {Bingying}\ \bibnamefont {Pan}}, \bibinfo {author}
  {\bibfnamefont {Qisi}\ \bibnamefont {Wang}}, \bibinfo {author} {\bibfnamefont
  {H.~C.}\ \bibnamefont {Walker}}, \bibinfo {author} {\bibfnamefont
  {P.}~\bibnamefont {Steffens}}, \bibinfo {author} {\bibfnamefont
  {M.}~\bibnamefont {Boehm}}, \bibinfo {author} {\bibfnamefont {Yiqing}\
  \bibnamefont {Hao}}, \bibinfo {author} {\bibfnamefont {D.~L.}\ \bibnamefont
  {Quintero-Castro}}, \bibinfo {author} {\bibfnamefont {L.~W.}\ \bibnamefont
  {Harriger}}, \bibinfo {author} {\bibfnamefont {M.~D.}\ \bibnamefont
  {Frontzek}}, \bibinfo {author} {\bibfnamefont {Lijie}\ \bibnamefont {Hao}},
  \bibinfo {author} {\bibfnamefont {Siqin}\ \bibnamefont {Meng}}, \bibinfo
  {author} {\bibfnamefont {Qingming}\ \bibnamefont {Zhang}}, \bibinfo {author}
  {\bibfnamefont {Gang}\ \bibnamefont {Chen}}, \ and\ \bibinfo {author}
  {\bibfnamefont {Jun}\ \bibnamefont {Zhao}},\ }\bibfield  {title} {\enquote
  {\bibinfo {title} {{Evidence for a spinon Fermi surface in a
  triangular-lattice quantum-spin-liquid candidate}},}\ }\href@noop {}
  {\bibfield  {journal} {\bibinfo  {journal} {Nature}\ }\textbf {\bibinfo
  {volume} {540}},\ \bibinfo {pages} {559} (\bibinfo {year}
  {2016})}\BibitemShut {NoStop}%
\bibitem [{\citenamefont {Paddison}\ \emph {et~al.}(2017)\citenamefont
  {Paddison}, \citenamefont {Daum}, \citenamefont {Dun}, \citenamefont
  {Ehlers}, \citenamefont {Liu}, \citenamefont {Stone}, \citenamefont {Zhou},\
  and\ \citenamefont {Mourigal}}]{np13_117}%
  \BibitemOpen
  \bibfield  {author} {\bibinfo {author} {\bibfnamefont {Joseph A.~M.}\
  \bibnamefont {Paddison}}, \bibinfo {author} {\bibfnamefont {Marcus}\
  \bibnamefont {Daum}}, \bibinfo {author} {\bibfnamefont {Zhiling}\
  \bibnamefont {Dun}}, \bibinfo {author} {\bibfnamefont {Georg}\ \bibnamefont
  {Ehlers}}, \bibinfo {author} {\bibfnamefont {Yaohua}\ \bibnamefont {Liu}},
  \bibinfo {author} {\bibfnamefont {Matthew~B.}\ \bibnamefont {Stone}},
  \bibinfo {author} {\bibfnamefont {Haidong}\ \bibnamefont {Zhou}}, \ and\
  \bibinfo {author} {\bibfnamefont {Martin}\ \bibnamefont {Mourigal}},\
  }\bibfield  {title} {\enquote {\bibinfo {title} {{Continuous excitations of
  the triangular-lattice quantum spin liquid YbMgGaO$_4$}},}\ }\href@noop {}
  {\bibfield  {journal} {\bibinfo  {journal} {Nat. Phys.}\ }\textbf {\bibinfo
  {volume} {13}},\ \bibinfo {pages} {117} (\bibinfo {year}
  {2017})}\BibitemShut {NoStop}%
\bibitem [{\citenamefont {Li}\ \emph {et~al.}(2016{\natexlab{b}})\citenamefont
  {Li}, \citenamefont {Adroja}, \citenamefont {Biswas}, \citenamefont {Baker},
  \citenamefont {Zhang}, \citenamefont {Liu}, \citenamefont {Tsirlin},
  \citenamefont {Gegenwart},\ and\ \citenamefont
  {Zhang}}]{PhysRevLett.117.097201}%
  \BibitemOpen
  \bibfield  {author} {\bibinfo {author} {\bibfnamefont {Yuesheng}\
  \bibnamefont {Li}}, \bibinfo {author} {\bibfnamefont {Devashibhai}\
  \bibnamefont {Adroja}}, \bibinfo {author} {\bibfnamefont {Pabitra~K.}\
  \bibnamefont {Biswas}}, \bibinfo {author} {\bibfnamefont {Peter~J.}\
  \bibnamefont {Baker}}, \bibinfo {author} {\bibfnamefont {Qian}\ \bibnamefont
  {Zhang}}, \bibinfo {author} {\bibfnamefont {Juanjuan}\ \bibnamefont {Liu}},
  \bibinfo {author} {\bibfnamefont {Alexander~A.}\ \bibnamefont {Tsirlin}},
  \bibinfo {author} {\bibfnamefont {Philipp}\ \bibnamefont {Gegenwart}}, \ and\
  \bibinfo {author} {\bibfnamefont {Qingming}\ \bibnamefont {Zhang}},\
  }\bibfield  {title} {\enquote {\bibinfo {title} {{Muon Spin Relaxation
  Evidence for the U(1) Quantum Spin-Liquid Ground State in the Triangular
  Antiferromagnet YbMgGaO$_{4}$}},}\ }\href {\doibase
  10.1103/PhysRevLett.117.097201} {\bibfield  {journal} {\bibinfo  {journal}
  {Phys. Rev. Lett.}\ }\textbf {\bibinfo {volume} {117}},\ \bibinfo {pages}
  {097201} (\bibinfo {year} {2016}{\natexlab{b}})}\BibitemShut {NoStop}%
\bibitem [{\citenamefont {Li}\ \emph {et~al.}(2017{\natexlab{a}})\citenamefont
  {Li}, \citenamefont {Adroja}, \citenamefont {Bewley}, \citenamefont
  {Voneshen}, \citenamefont {Tsirlin}, \citenamefont {Gegenwart},\ and\
  \citenamefont {Zhang}}]{PhysRevLett.118.107202}%
  \BibitemOpen
  \bibfield  {author} {\bibinfo {author} {\bibfnamefont {Yuesheng}\
  \bibnamefont {Li}}, \bibinfo {author} {\bibfnamefont {Devashibhai}\
  \bibnamefont {Adroja}}, \bibinfo {author} {\bibfnamefont {Robert~I.}\
  \bibnamefont {Bewley}}, \bibinfo {author} {\bibfnamefont {David}\
  \bibnamefont {Voneshen}}, \bibinfo {author} {\bibfnamefont {Alexander~A.}\
  \bibnamefont {Tsirlin}}, \bibinfo {author} {\bibfnamefont {Philipp}\
  \bibnamefont {Gegenwart}}, \ and\ \bibinfo {author} {\bibfnamefont
  {Qingming}\ \bibnamefont {Zhang}},\ }\bibfield  {title} {\enquote {\bibinfo
  {title} {{Crystalline Electric-Field Randomness in the Triangular Lattice
  Spin-Liquid ${\mathrm{YbMgGaO}}_{4}$}},}\ }\href {\doibase
  10.1103/PhysRevLett.118.107202} {\bibfield  {journal} {\bibinfo  {journal}
  {Phys. Rev. Lett.}\ }\textbf {\bibinfo {volume} {118}},\ \bibinfo {pages}
  {107202} (\bibinfo {year} {2017}{\natexlab{a}})}\BibitemShut {NoStop}%
\bibitem [{\citenamefont {Li}\ \emph {et~al.}(2017{\natexlab{b}})\citenamefont
  {Li}, \citenamefont {Adroja}, \citenamefont {Voneshen}, \citenamefont
  {Bewley}, \citenamefont {Zhang}, \citenamefont {Tsirlin},\ and\ \citenamefont
  {Gegenwart}}]{nc8_15814}%
  \BibitemOpen
  \bibfield  {author} {\bibinfo {author} {\bibfnamefont {Yuesheng}\
  \bibnamefont {Li}}, \bibinfo {author} {\bibfnamefont {Devashibhai}\
  \bibnamefont {Adroja}}, \bibinfo {author} {\bibfnamefont {David}\
  \bibnamefont {Voneshen}}, \bibinfo {author} {\bibfnamefont {Robert~I.}\
  \bibnamefont {Bewley}}, \bibinfo {author} {\bibfnamefont {Qingming}\
  \bibnamefont {Zhang}}, \bibinfo {author} {\bibfnamefont {Alexander~A.}\
  \bibnamefont {Tsirlin}}, \ and\ \bibinfo {author} {\bibfnamefont {Philipp}\
  \bibnamefont {Gegenwart}},\ }\bibfield  {title} {\enquote {\bibinfo {title}
  {{Nearest-neighbour resonating valence bonds in YbMgGaO$_4$}},}\ }\href@noop
  {} {\bibfield  {journal} {\bibinfo  {journal} {Nat. Commun.}\ }\textbf
  {\bibinfo {volume} {8}},\ \bibinfo {pages} {15814} (\bibinfo {year}
  {2017}{\natexlab{b}})}\BibitemShut {NoStop}%
\bibitem [{\citenamefont {Li}\ \emph {et~al.}(2019)\citenamefont {Li},
  \citenamefont {Bachus}, \citenamefont {Liu}, \citenamefont {Radelytskyi},
  \citenamefont {Bertin}, \citenamefont {Schneidewind}, \citenamefont {Tokiwa},
  \citenamefont {Tsirlin},\ and\ \citenamefont
  {Gegenwart}}]{PhysRevLett.122.137201}%
  \BibitemOpen
  \bibfield  {author} {\bibinfo {author} {\bibfnamefont {Yuesheng}\
  \bibnamefont {Li}}, \bibinfo {author} {\bibfnamefont {Sebastian}\
  \bibnamefont {Bachus}}, \bibinfo {author} {\bibfnamefont {Benqiong}\
  \bibnamefont {Liu}}, \bibinfo {author} {\bibfnamefont {Igor}\ \bibnamefont
  {Radelytskyi}}, \bibinfo {author} {\bibfnamefont {Alexandre}\ \bibnamefont
  {Bertin}}, \bibinfo {author} {\bibfnamefont {Astrid}\ \bibnamefont
  {Schneidewind}}, \bibinfo {author} {\bibfnamefont {Yoshifumi}\ \bibnamefont
  {Tokiwa}}, \bibinfo {author} {\bibfnamefont {Alexander~A.}\ \bibnamefont
  {Tsirlin}}, \ and\ \bibinfo {author} {\bibfnamefont {Philipp}\ \bibnamefont
  {Gegenwart}},\ }\bibfield  {title} {\enquote {\bibinfo {title}
  {{Rearrangement of Uncorrelated Valence Bonds Evidenced by Low-Energy Spin
  Excitations in ${\mathrm{YbMgGaO}}_{4}$}},}\ }\href {\doibase
  10.1103/PhysRevLett.122.137201} {\bibfield  {journal} {\bibinfo  {journal}
  {Phys. Rev. Lett.}\ }\textbf {\bibinfo {volume} {122}},\ \bibinfo {pages}
  {137201} (\bibinfo {year} {2019})}\BibitemShut {NoStop}%
\bibitem [{\citenamefont {Xu}\ \emph {et~al.}(2016)\citenamefont {Xu},
  \citenamefont {Zhang}, \citenamefont {Li}, \citenamefont {Yu}, \citenamefont
  {Hong}, \citenamefont {Zhang},\ and\ \citenamefont
  {Li}}]{PhysRevLett.117.267202}%
  \BibitemOpen
  \bibfield  {author} {\bibinfo {author} {\bibfnamefont {Y.}~\bibnamefont
  {Xu}}, \bibinfo {author} {\bibfnamefont {J.}~\bibnamefont {Zhang}}, \bibinfo
  {author} {\bibfnamefont {Y.~S.}\ \bibnamefont {Li}}, \bibinfo {author}
  {\bibfnamefont {Y.~J.}\ \bibnamefont {Yu}}, \bibinfo {author} {\bibfnamefont
  {X.~C.}\ \bibnamefont {Hong}}, \bibinfo {author} {\bibfnamefont {Q.~M.}\
  \bibnamefont {Zhang}}, \ and\ \bibinfo {author} {\bibfnamefont {S.~Y.}\
  \bibnamefont {Li}},\ }\bibfield  {title} {\enquote {\bibinfo {title}
  {{Absence of Magnetic Thermal Conductivity in the Quantum Spin-Liquid
  Candidate ${\mathrm{YbMgGaO}}_{4}$}},}\ }\href {\doibase
  10.1103/PhysRevLett.117.267202} {\bibfield  {journal} {\bibinfo  {journal}
  {Phys. Rev. Lett.}\ }\textbf {\bibinfo {volume} {117}},\ \bibinfo {pages}
  {267202} (\bibinfo {year} {2016})}\BibitemShut {NoStop}%
\bibitem [{\citenamefont {Ma}\ \emph {et~al.}(2018{\natexlab{a}})\citenamefont
  {Ma}, \citenamefont {Wang}, \citenamefont {Dong}, \citenamefont {Zhang},
  \citenamefont {Li}, \citenamefont {Zheng}, \citenamefont {Yu}, \citenamefont
  {Wang}, \citenamefont {Che}, \citenamefont {Ran}, \citenamefont {Bao},
  \citenamefont {Cai}, \citenamefont {\ifmmode~\check{C}\else
  \v{C}\fi{}erm\'ak}, \citenamefont {Schneidewind}, \citenamefont {Yano},
  \citenamefont {Gardner}, \citenamefont {Lu}, \citenamefont {Yu},
  \citenamefont {Liu}, \citenamefont {Li}, \citenamefont {Li},\ and\
  \citenamefont {Wen}}]{PhysRevLett.120.087201}%
  \BibitemOpen
  \bibfield  {author} {\bibinfo {author} {\bibfnamefont {Zhen}\ \bibnamefont
  {Ma}}, \bibinfo {author} {\bibfnamefont {Jinghui}\ \bibnamefont {Wang}},
  \bibinfo {author} {\bibfnamefont {Zhao-Yang}\ \bibnamefont {Dong}}, \bibinfo
  {author} {\bibfnamefont {Jun}\ \bibnamefont {Zhang}}, \bibinfo {author}
  {\bibfnamefont {Shichao}\ \bibnamefont {Li}}, \bibinfo {author}
  {\bibfnamefont {Shu-Han}\ \bibnamefont {Zheng}}, \bibinfo {author}
  {\bibfnamefont {Yunjie}\ \bibnamefont {Yu}}, \bibinfo {author} {\bibfnamefont
  {Wei}\ \bibnamefont {Wang}}, \bibinfo {author} {\bibfnamefont {Liqiang}\
  \bibnamefont {Che}}, \bibinfo {author} {\bibfnamefont {Kejing}\ \bibnamefont
  {Ran}}, \bibinfo {author} {\bibfnamefont {Song}\ \bibnamefont {Bao}},
  \bibinfo {author} {\bibfnamefont {Zhengwei}\ \bibnamefont {Cai}}, \bibinfo
  {author} {\bibfnamefont {P.}~\bibnamefont {\ifmmode~\check{C}\else
  \v{C}\fi{}erm\'ak}}, \bibinfo {author} {\bibfnamefont {A.}~\bibnamefont
  {Schneidewind}}, \bibinfo {author} {\bibfnamefont {S.}~\bibnamefont {Yano}},
  \bibinfo {author} {\bibfnamefont {J.~S.}\ \bibnamefont {Gardner}}, \bibinfo
  {author} {\bibfnamefont {Xin}\ \bibnamefont {Lu}}, \bibinfo {author}
  {\bibfnamefont {Shun-Li}\ \bibnamefont {Yu}}, \bibinfo {author}
  {\bibfnamefont {Jun-Ming}\ \bibnamefont {Liu}}, \bibinfo {author}
  {\bibfnamefont {Shiyan}\ \bibnamefont {Li}}, \bibinfo {author} {\bibfnamefont
  {Jian-Xin}\ \bibnamefont {Li}}, \ and\ \bibinfo {author} {\bibfnamefont
  {Jinsheng}\ \bibnamefont {Wen}},\ }\bibfield  {title} {\enquote {\bibinfo
  {title} {{Spin-Glass Ground State in a Triangular-Lattice Compound
  ${\mathrm{YbZnGaO}}_{4}$}},}\ }\href {\doibase
  10.1103/PhysRevLett.120.087201} {\bibfield  {journal} {\bibinfo  {journal}
  {Phys. Rev. Lett.}\ }\textbf {\bibinfo {volume} {120}},\ \bibinfo {pages}
  {087201} (\bibinfo {year} {2018}{\natexlab{a}})}\BibitemShut {NoStop}%
\bibitem [{\citenamefont {Ma}\ \emph {et~al.}(2018{\natexlab{b}})\citenamefont
  {Ma}, \citenamefont {Ran}, \citenamefont {Wang}, \citenamefont {Bao},
  \citenamefont {Cai}, \citenamefont {Li},\ and\ \citenamefont
  {Wen}}]{ZhenMa:106101}%
  \BibitemOpen
  \bibfield  {author} {\bibinfo {author} {\bibfnamefont {Zhen}\ \bibnamefont
  {Ma}}, \bibinfo {author} {\bibfnamefont {Kejing}\ \bibnamefont {Ran}},
  \bibinfo {author} {\bibfnamefont {Jinghui}\ \bibnamefont {Wang}}, \bibinfo
  {author} {\bibfnamefont {Song}\ \bibnamefont {Bao}}, \bibinfo {author}
  {\bibfnamefont {Zhengwei}\ \bibnamefont {Cai}}, \bibinfo {author}
  {\bibfnamefont {Shichao}\ \bibnamefont {Li}}, \ and\ \bibinfo {author}
  {\bibfnamefont {Jinsheng}\ \bibnamefont {Wen}},\ }\bibfield  {title}
  {\enquote {\bibinfo {title} {Recent progress on magnetic-field studies on
  quantum-spin-liquid candidates},}\ }\href {\doibase
  10.1088/1674-1056/27/10/106101} {\bibfield  {journal} {\bibinfo  {journal}
  {Chin. Phys. B}\ }\textbf {\bibinfo {volume} {27}},\ \bibinfo {eid}
  {106101} (\bibinfo {year} {2018}{\natexlab{b}})}\BibitemShut {NoStop}%
\bibitem [{\citenamefont {Kimchi}\ \emph {et~al.}(2018)\citenamefont {Kimchi},
  \citenamefont {Nahum},\ and\ \citenamefont {Senthil}}]{PhysRevX.8.031028}%
  \BibitemOpen
  \bibfield  {author} {\bibinfo {author} {\bibfnamefont {Itamar}\ \bibnamefont
  {Kimchi}}, \bibinfo {author} {\bibfnamefont {Adam}\ \bibnamefont {Nahum}}, \
  and\ \bibinfo {author} {\bibfnamefont {T.}~\bibnamefont {Senthil}},\
  }\bibfield  {title} {\enquote {\bibinfo {title} {{Valence Bonds in Random
  Quantum Magnets: Theory and Application to ${\mathrm{YbMgGaO}}_{4}$}},}\
  }\href {\doibase 10.1103/PhysRevX.8.031028} {\bibfield  {journal} {\bibinfo
  {journal} {Phys. Rev. X}\ }\textbf {\bibinfo {volume} {8}},\ \bibinfo {pages}
  {031028} (\bibinfo {year} {2018})}\BibitemShut {NoStop}%
\bibitem [{\citenamefont {Wen}\ \emph {et~al.}(2019)\citenamefont {Wen},
  \citenamefont {Yu}, \citenamefont {Li}, \citenamefont {Yu},\ and\
  \citenamefont {Li}}]{npjqm4_12}%
  \BibitemOpen
  \bibfield  {author} {\bibinfo {author} {\bibfnamefont {Jinsheng}\
  \bibnamefont {Wen}}, \bibinfo {author} {\bibfnamefont {Shun-Li}\ \bibnamefont
  {Yu}}, \bibinfo {author} {\bibfnamefont {Shiyan}\ \bibnamefont {Li}},
  \bibinfo {author} {\bibfnamefont {Weiqiang}\ \bibnamefont {Yu}}, \ and\
  \bibinfo {author} {\bibfnamefont {Jian-Xin}\ \bibnamefont {Li}},\ }\bibfield
  {title} {\enquote {\bibinfo {title} {Experimental identification of quantum
  spin liquids},}\ }\href@noop {} {\bibfield  {journal} {\bibinfo  {journal}
  {npj Quant. Mater.}\ }\textbf {\bibinfo {volume} {4}},\ \bibinfo {pages} {12}
  (\bibinfo {year} {2019})}\BibitemShut {NoStop}%
\bibitem [{\citenamefont {Zhu}\ \emph {et~al.}(2017)\citenamefont {Zhu},
  \citenamefont {Maksimov}, \citenamefont {White},\ and\ \citenamefont
  {Chernyshev}}]{PhysRevLett.119.157201}%
  \BibitemOpen
  \bibfield  {author} {\bibinfo {author} {\bibfnamefont {Zhenyue}\ \bibnamefont
  {Zhu}}, \bibinfo {author} {\bibfnamefont {P.~A.}\ \bibnamefont {Maksimov}},
  \bibinfo {author} {\bibfnamefont {Steven~R.}\ \bibnamefont {White}}, \ and\
  \bibinfo {author} {\bibfnamefont {A.~L.}\ \bibnamefont {Chernyshev}},\
  }\bibfield  {title} {\enquote {\bibinfo {title} {{Disorder-Induced Mimicry of
  a Spin Liquid in ${\mathrm{YbMgGaO}}_{4}$}},}\ }\href {\doibase
  10.1103/PhysRevLett.119.157201} {\bibfield  {journal} {\bibinfo  {journal}
  {Phys. Rev. Lett.}\ }\textbf {\bibinfo {volume} {119}},\ \bibinfo {pages}
  {157201} (\bibinfo {year} {2017})}\BibitemShut {NoStop}%
\bibitem [{\citenamefont {Ma}\ \emph {et~al.}(2020)\citenamefont {Ma},
  \citenamefont {Dong}, \citenamefont {Wu}, \citenamefont {Zhu}, \citenamefont
  {Bao}, \citenamefont {Cai}, \citenamefont {Wang}, \citenamefont {Shangguan},
  \citenamefont {Wang}, \citenamefont {Ran}, \citenamefont {Yu}, \citenamefont
  {Deng}, \citenamefont {Mole}, \citenamefont {Li}, \citenamefont {Yu},
  \citenamefont {Li},\ and\ \citenamefont {Wen}}]{PhysRevB.102.224415}%
  \BibitemOpen
  \bibfield  {author} {\bibinfo {author} {\bibfnamefont {Zhen}\ \bibnamefont
  {Ma}}, \bibinfo {author} {\bibfnamefont {Zhao-Yang}\ \bibnamefont {Dong}},
  \bibinfo {author} {\bibfnamefont {Si}~\bibnamefont {Wu}}, \bibinfo {author}
  {\bibfnamefont {Yinghao}\ \bibnamefont {Zhu}}, \bibinfo {author}
  {\bibfnamefont {Song}\ \bibnamefont {Bao}}, \bibinfo {author} {\bibfnamefont
  {Zhengwei}\ \bibnamefont {Cai}}, \bibinfo {author} {\bibfnamefont {Wei}\
  \bibnamefont {Wang}}, \bibinfo {author} {\bibfnamefont {Yanyan}\ \bibnamefont
  {Shangguan}}, \bibinfo {author} {\bibfnamefont {Jinghui}\ \bibnamefont
  {Wang}}, \bibinfo {author} {\bibfnamefont {Kejing}\ \bibnamefont {Ran}},
  \bibinfo {author} {\bibfnamefont {Dehong}\ \bibnamefont {Yu}}, \bibinfo
  {author} {\bibfnamefont {Guochu}\ \bibnamefont {Deng}}, \bibinfo {author}
  {\bibfnamefont {Richard~A.}\ \bibnamefont {Mole}}, \bibinfo {author}
  {\bibfnamefont {Hai-Feng}\ \bibnamefont {Li}}, \bibinfo {author}
  {\bibfnamefont {Shun-Li}\ \bibnamefont {Yu}}, \bibinfo {author}
  {\bibfnamefont {Jian-Xin}\ \bibnamefont {Li}}, \ and\ \bibinfo {author}
  {\bibfnamefont {Jinsheng}\ \bibnamefont {Wen}},\ }\bibfield  {title}
  {\enquote {\bibinfo {title} {Disorder-induced spin-liquid-like behavior in
  kagome-lattice compounds},}\ }\href {\doibase 10.1103/PhysRevB.102.224415}
  {\bibfield  {journal} {\bibinfo  {journal} {Phys. Rev. B}\ }\textbf {\bibinfo
  {volume} {102}},\ \bibinfo {pages} {224415} (\bibinfo {year}
  {2020})}\BibitemShut {NoStop}%
\bibitem [{\citenamefont {Li}(2019)}]{https://doi.org/10.1002/qute.201900089}%
  \BibitemOpen
  \bibfield  {author} {\bibinfo {author} {\bibfnamefont {Yuesheng}\
  \bibnamefont {Li}},\ }\bibfield  {title} {\enquote {\bibinfo {title}
  {{YbMgGaO$_4$: A Triangular-Lattice Quantum Spin Liquid Candidate}},}\ }\href
  {\doibase https://doi.org/10.1002/qute.201900089} {\bibfield  {journal}
  {\bibinfo  {journal} {Adv. Quant. Technol.}\ }\textbf {\bibinfo
  {volume} {2}},\ \bibinfo {pages} {1900089} (\bibinfo {year} {2019})}\BibitemShut
  {NoStop}%
\bibitem [{\citenamefont {Liu}\ \emph {et~al.}(2022{\natexlab{a}})\citenamefont
  {Liu}, \citenamefont {Yuan}, \citenamefont {Li}, \citenamefont {Li},
  \citenamefont {Zhao}, \citenamefont {Liao},\ and\ \citenamefont
  {Li}}]{PhysRevB.105.024418}%
  \BibitemOpen
  \bibfield  {author} {\bibinfo {author} {\bibfnamefont {Jiabin}\ \bibnamefont
  {Liu}}, \bibinfo {author} {\bibfnamefont {Long}\ \bibnamefont {Yuan}},
  \bibinfo {author} {\bibfnamefont {Xuan}\ \bibnamefont {Li}}, \bibinfo
  {author} {\bibfnamefont {Boqiang}\ \bibnamefont {Li}}, \bibinfo {author}
  {\bibfnamefont {Kan}\ \bibnamefont {Zhao}}, \bibinfo {author} {\bibfnamefont
  {Haijun}\ \bibnamefont {Liao}}, \ and\ \bibinfo {author} {\bibfnamefont
  {Yuesheng}\ \bibnamefont {Li}},\ }\bibfield  {title} {\enquote {\bibinfo
  {title} {{Gapless spin liquid behavior in a kagome Heisenberg antiferromagnet
  with randomly distributed hexagons of alternate bonds}},}\ }\href {\doibase
  10.1103/PhysRevB.105.024418} {\bibfield  {journal} {\bibinfo  {journal}
  {Phys. Rev. B}\ }\textbf {\bibinfo {volume} {105}},\ \bibinfo {pages}
  {024418} (\bibinfo {year} {2022}{\natexlab{a}})}\BibitemShut {NoStop}%
\bibitem [{\citenamefont {Furukawa}\ \emph {et~al.}(2015)\citenamefont
  {Furukawa}, \citenamefont {Miyagawa}, \citenamefont {Itou}, \citenamefont
  {Ito}, \citenamefont {Taniguchi}, \citenamefont {Saito}, \citenamefont
  {Iguchi}, \citenamefont {Sasaki},\ and\ \citenamefont
  {Kanoda}}]{PhysRevLett.115.077001}%
  \BibitemOpen
  \bibfield  {author} {\bibinfo {author} {\bibfnamefont {T.}~\bibnamefont
  {Furukawa}}, \bibinfo {author} {\bibfnamefont {K.}~\bibnamefont {Miyagawa}},
  \bibinfo {author} {\bibfnamefont {T.}~\bibnamefont {Itou}}, \bibinfo {author}
  {\bibfnamefont {M.}~\bibnamefont {Ito}}, \bibinfo {author} {\bibfnamefont
  {H.}~\bibnamefont {Taniguchi}}, \bibinfo {author} {\bibfnamefont
  {M.}~\bibnamefont {Saito}}, \bibinfo {author} {\bibfnamefont
  {S.}~\bibnamefont {Iguchi}}, \bibinfo {author} {\bibfnamefont
  {T.}~\bibnamefont {Sasaki}}, \ and\ \bibinfo {author} {\bibfnamefont
  {K.}~\bibnamefont {Kanoda}},\ }\bibfield  {title} {\enquote {\bibinfo {title}
  {{Quantum Spin Liquid Emerging from Antiferromagnetic Order by Introducing
  Disorder}},}\ }\href {\doibase 10.1103/PhysRevLett.115.077001} {\bibfield
  {journal} {\bibinfo  {journal} {Phys. Rev. Lett.}\ }\textbf {\bibinfo
  {volume} {115}},\ \bibinfo {pages} {077001} (\bibinfo {year}
  {2015})}\BibitemShut {NoStop}%
\bibitem [{\citenamefont {Hu}\ \emph {et~al.}(2021)\citenamefont {Hu},
  \citenamefont {Pajerowski}, \citenamefont {Zhang}, \citenamefont
  {Podlesnyak}, \citenamefont {Qiu}, \citenamefont {Huang}, \citenamefont
  {Zhou}, \citenamefont {Klich}, \citenamefont {Kolesnikov}, \citenamefont
  {Stone},\ and\ \citenamefont {Lee}}]{PhysRevLett.127.017201}%
  \BibitemOpen
  \bibfield  {author} {\bibinfo {author} {\bibfnamefont {Xiao}\ \bibnamefont
  {Hu}}, \bibinfo {author} {\bibfnamefont {Daniel~M.}\ \bibnamefont
  {Pajerowski}}, \bibinfo {author} {\bibfnamefont {Depei}\ \bibnamefont
  {Zhang}}, \bibinfo {author} {\bibfnamefont {Andrey~A.}\ \bibnamefont
  {Podlesnyak}}, \bibinfo {author} {\bibfnamefont {Yiming}\ \bibnamefont
  {Qiu}}, \bibinfo {author} {\bibfnamefont {Qing}\ \bibnamefont {Huang}},
  \bibinfo {author} {\bibfnamefont {Haidong}\ \bibnamefont {Zhou}}, \bibinfo
  {author} {\bibfnamefont {Israel}\ \bibnamefont {Klich}}, \bibinfo {author}
  {\bibfnamefont {Alexander~I.}\ \bibnamefont {Kolesnikov}}, \bibinfo {author}
  {\bibfnamefont {Matthew~B.}\ \bibnamefont {Stone}}, \ and\ \bibinfo {author}
  {\bibfnamefont {Seung-Hun}\ \bibnamefont {Lee}},\ }\bibfield  {title}
  {\enquote {\bibinfo {title} {{Freezing of a Disorder Induced Spin Liquid with
  Strong Quantum Fluctuations}},}\ }\href {\doibase
  10.1103/PhysRevLett.127.017201} {\bibfield  {journal} {\bibinfo  {journal}
  {Phys. Rev. Lett.}\ }\textbf {\bibinfo {volume} {127}},\ \bibinfo {pages}
  {017201} (\bibinfo {year} {2021})}\BibitemShut {NoStop}%
\bibitem [{\citenamefont {M.~Bordelon}\ \emph {et~al.}(2019)\citenamefont
  {M.~Bordelon}, \citenamefont {Kenney}, \citenamefont {Liu}, \citenamefont
  {Hogan}, \citenamefont {Posthuma}, \citenamefont {Kavand}, \citenamefont
  {Lyu}, \citenamefont {Sherwin}, \citenamefont {Butch}, \citenamefont {Brown},
  \citenamefont {Graf}, \citenamefont {Balents},\ and\ \citenamefont
  {Wilson}}]{np15_1058}%
  \BibitemOpen
  \bibfield  {author} {\bibinfo {author} {\bibfnamefont {Mitchell}\
  \bibnamefont {M.~Bordelon}}, \bibinfo {author} {\bibfnamefont {Eric}\
  \bibnamefont {Kenney}}, \bibinfo {author} {\bibfnamefont {Chunxiao}\
  \bibnamefont {Liu}}, \bibinfo {author} {\bibfnamefont {Tom}\ \bibnamefont
  {Hogan}}, \bibinfo {author} {\bibfnamefont {Lorenzo}\ \bibnamefont
  {Posthuma}}, \bibinfo {author} {\bibfnamefont {Marzieh}\ \bibnamefont
  {Kavand}}, \bibinfo {author} {\bibfnamefont {Yuanqi}\ \bibnamefont {Lyu}},
  \bibinfo {author} {\bibfnamefont {Mark}\ \bibnamefont {Sherwin}}, \bibinfo
  {author} {\bibfnamefont {N.~P.}\ \bibnamefont {Butch}}, \bibinfo {author}
  {\bibfnamefont {Craig}\ \bibnamefont {Brown}}, \bibinfo {author}
  {\bibfnamefont {M.~J.}\ \bibnamefont {Graf}}, \bibinfo {author}
  {\bibfnamefont {Leon}\ \bibnamefont {Balents}}, \ and\ \bibinfo {author}
  {\bibfnamefont {Stephen~D.}\ \bibnamefont {Wilson}},\ }\bibfield  {title}
  {\enquote {\bibinfo {title} {{Field-tunable quantum disordered ground state
  in the triangular lattice antiferromagnet NaYbO$_2$}},}\ }\href@noop {}
  {\bibfield  {journal} {\bibinfo  {journal} {Nat. Phys.}\ }\textbf {\bibinfo
  {volume} {15}},\ \bibinfo {pages} {1058} (\bibinfo {year}
  {2019})}\BibitemShut {NoStop}%
\bibitem [{\citenamefont {Yan}\ \emph {et~al.}(2019)\citenamefont {Yan},
  \citenamefont {Okamoto}, \citenamefont {Wu}, \citenamefont {Zheng},
  \citenamefont {Zhou}, \citenamefont {Cao},\ and\ \citenamefont
  {McGuire}}]{PhysRevMaterials.3.074405}%
  \BibitemOpen
  \bibfield  {author} {\bibinfo {author} {\bibfnamefont {J.-Q.}\ \bibnamefont
  {Yan}}, \bibinfo {author} {\bibfnamefont {S.}~\bibnamefont {Okamoto}},
  \bibinfo {author} {\bibfnamefont {Y.}~\bibnamefont {Wu}}, \bibinfo {author}
  {\bibfnamefont {Q.}~\bibnamefont {Zheng}}, \bibinfo {author} {\bibfnamefont
  {H.~D.}\ \bibnamefont {Zhou}}, \bibinfo {author} {\bibfnamefont {H.~B.}\
  \bibnamefont {Cao}}, \ and\ \bibinfo {author} {\bibfnamefont {M.~A.}\
  \bibnamefont {McGuire}},\ }\bibfield  {title} {\enquote {\bibinfo {title}
  {{Magnetic order in single crystals of
  ${\mathrm{Na}}_{3}{\mathrm{Co}}_{2}{\mathrm{SbO}}_{6}$ with a honeycomb
  arrangement of 3d$^7$ Co$^{2+}$ ions}},}\ }\href {\doibase
  10.1103/PhysRevMaterials.3.074405} {\bibfield  {journal} {\bibinfo  {journal}
  {Phys. Rev. Mater.}\ }\textbf {\bibinfo {volume} {3}},\ \bibinfo {pages}
  {074405} (\bibinfo {year} {2019})}\BibitemShut {NoStop}%
\bibitem [{\citenamefont {Dai}\ \emph {et~al.}(2021)\citenamefont {Dai},
  \citenamefont {Zhang}, \citenamefont {Xie}, \citenamefont {Duan},
  \citenamefont {Gao}, \citenamefont {Zhu}, \citenamefont {Feng}, \citenamefont
  {Tao}, \citenamefont {Huang}, \citenamefont {Cao}, \citenamefont
  {Podlesnyak}, \citenamefont {Granroth}, \citenamefont {Everett},
  \citenamefont {Neuefeind}, \citenamefont {Voneshen}, \citenamefont {Wang},
  \citenamefont {Tan}, \citenamefont {Morosan}, \citenamefont {Wang},
  \citenamefont {Lin}, \citenamefont {Shu}, \citenamefont {Chen}, \citenamefont
  {Guo}, \citenamefont {Lu},\ and\ \citenamefont {Dai}}]{PhysRevX.11.021044}%
  \BibitemOpen
  \bibfield  {author} {\bibinfo {author} {\bibfnamefont {Peng-Ling}\
  \bibnamefont {Dai}}, \bibinfo {author} {\bibfnamefont {Gaoning}\ \bibnamefont
  {Zhang}}, \bibinfo {author} {\bibfnamefont {Yaofeng}\ \bibnamefont {Xie}},
  \bibinfo {author} {\bibfnamefont {Chunruo}\ \bibnamefont {Duan}}, \bibinfo
  {author} {\bibfnamefont {Yonghao}\ \bibnamefont {Gao}}, \bibinfo {author}
  {\bibfnamefont {Zihao}\ \bibnamefont {Zhu}}, \bibinfo {author} {\bibfnamefont
  {Erxi}\ \bibnamefont {Feng}}, \bibinfo {author} {\bibfnamefont {Zhen}\
  \bibnamefont {Tao}}, \bibinfo {author} {\bibfnamefont {Chien-Lung}\
  \bibnamefont {Huang}}, \bibinfo {author} {\bibfnamefont {Huibo}\ \bibnamefont
  {Cao}}, \bibinfo {author} {\bibfnamefont {Andrey}\ \bibnamefont
  {Podlesnyak}}, \bibinfo {author} {\bibfnamefont {Garrett~E.}\ \bibnamefont
  {Granroth}}, \bibinfo {author} {\bibfnamefont {Michelle~S.}\ \bibnamefont
  {Everett}}, \bibinfo {author} {\bibfnamefont {Joerg~C.}\ \bibnamefont
  {Neuefeind}}, \bibinfo {author} {\bibfnamefont {David}\ \bibnamefont
  {Voneshen}}, \bibinfo {author} {\bibfnamefont {Shun}\ \bibnamefont {Wang}},
  \bibinfo {author} {\bibfnamefont {Guotai}\ \bibnamefont {Tan}}, \bibinfo
  {author} {\bibfnamefont {Emilia}\ \bibnamefont {Morosan}}, \bibinfo {author}
  {\bibfnamefont {Xia}\ \bibnamefont {Wang}}, \bibinfo {author} {\bibfnamefont
  {Hai-Qing}\ \bibnamefont {Lin}}, \bibinfo {author} {\bibfnamefont {Lei}\
  \bibnamefont {Shu}}, \bibinfo {author} {\bibfnamefont {Gang}\ \bibnamefont
  {Chen}}, \bibinfo {author} {\bibfnamefont {Yanfeng}\ \bibnamefont {Guo}},
  \bibinfo {author} {\bibfnamefont {Xingye}\ \bibnamefont {Lu}}, \ and\
  \bibinfo {author} {\bibfnamefont {Pengcheng}\ \bibnamefont {Dai}},\
  }\bibfield  {title} {\enquote {\bibinfo {title} {{Spinon Fermi Surface Spin
  Liquid in a Triangular Lattice Antiferromagnet ${\mathrm{NaYbSe}}_{2}$}},}\
  }\href {\doibase 10.1103/PhysRevX.11.021044} {\bibfield  {journal} {\bibinfo
  {journal} {Phys. Rev. X}\ }\textbf {\bibinfo {volume} {11}},\ \bibinfo
  {pages} {021044} (\bibinfo {year} {2021})}\BibitemShut {NoStop}%
\bibitem [{\citenamefont {Dufault}\ \emph {et~al.}(2023)\citenamefont
  {Dufault}, \citenamefont {Bahrami}, \citenamefont {Streeter}, \citenamefont
  {Yao}, \citenamefont {Gonzalez}, \citenamefont {Zhang},\ and\ \citenamefont
  {Tafti}}]{PhysRevB.108.064405}%
  \BibitemOpen
  \bibfield  {author} {\bibinfo {author} {\bibfnamefont {Emilie}\ \bibnamefont
  {Dufault}}, \bibinfo {author} {\bibfnamefont {Faranak}\ \bibnamefont
  {Bahrami}}, \bibinfo {author} {\bibfnamefont {Alenna}\ \bibnamefont
  {Streeter}}, \bibinfo {author} {\bibfnamefont {Xiaohan}\ \bibnamefont {Yao}},
  \bibinfo {author} {\bibfnamefont {Enrique}\ \bibnamefont {Gonzalez}},
  \bibinfo {author} {\bibfnamefont {Qiang}\ \bibnamefont {Zhang}}, \ and\
  \bibinfo {author} {\bibfnamefont {Fazel}\ \bibnamefont {Tafti}},\ }\bibfield
  {title} {\enquote {\bibinfo {title} {{Introducing the monoclinic polymorph of
  the honeycomb magnet
  ${\mathrm{Na}}_{2}{\mathrm{Co}}_{2}{\mathrm{TeO}}_{6}$}},}\ }\href {\doibase
  10.1103/PhysRevB.108.064405} {\bibfield  {journal} {\bibinfo  {journal}
  {Phys. Rev. B}\ }\textbf {\bibinfo {volume} {108}},\ \bibinfo {pages}
  {064405} (\bibinfo {year} {2023})}\BibitemShut {NoStop}%
\bibitem [{\citenamefont {Ashtar}\ \emph
  {et~al.}(2019{\natexlab{a}})\citenamefont {Ashtar}, \citenamefont {Gao},
  \citenamefont {Wang}, \citenamefont {Qiu}, \citenamefont {Tong},
  \citenamefont {Zou}, \citenamefont {Zhang}, \citenamefont {Marwat},
  \citenamefont {Yuan},\ and\ \citenamefont {Tian}}]{ASHTAR2019146}%
  \BibitemOpen
  \bibfield  {author} {\bibinfo {author} {\bibfnamefont {M.}~\bibnamefont
  {Ashtar}}, \bibinfo {author} {\bibfnamefont {Y.X.}\ \bibnamefont {Gao}},
  \bibinfo {author} {\bibfnamefont {C.L.}\ \bibnamefont {Wang}}, \bibinfo
  {author} {\bibfnamefont {Y.}~\bibnamefont {Qiu}}, \bibinfo {author}
  {\bibfnamefont {W.}~\bibnamefont {Tong}}, \bibinfo {author} {\bibfnamefont
  {Y.M.}\ \bibnamefont {Zou}}, \bibinfo {author} {\bibfnamefont {X.W.}\
  \bibnamefont {Zhang}}, \bibinfo {author} {\bibfnamefont {M.A.}\ \bibnamefont
  {Marwat}}, \bibinfo {author} {\bibfnamefont {S.L.}\ \bibnamefont {Yuan}}, \
  and\ \bibinfo {author} {\bibfnamefont {Z.M.}\ \bibnamefont {Tian}},\
  }\bibfield  {title} {\enquote {\bibinfo {title} {{Synthesis, structure and
  magnetic properties of rare-earth REMgAl$_{11}$O$_{19}$ (RE = Pr, Nd)
  compounds with two-dimensional triangular lattice}},}\ }\href {\doibase
  https://doi.org/10.1016/j.jallcom.2019.06.177} {\bibfield  {journal}
  {\bibinfo  {journal} {J. Alloy. Compd.}\ }\textbf {\bibinfo {volume} {802}},\
  \bibinfo {pages} {146} (\bibinfo {year}
  {2019}{\natexlab{a}})}\BibitemShut {NoStop}%
\bibitem [{\citenamefont {Ashtar}\ \emph
  {et~al.}(2019{\natexlab{b}})\citenamefont {Ashtar}, \citenamefont {Marwat},
  \citenamefont {Gao}, \citenamefont {Zhang}, \citenamefont {Pi}, \citenamefont
  {Yuan},\ and\ \citenamefont {Tian}}]{C9TC02643F}%
  \BibitemOpen
  \bibfield  {author} {\bibinfo {author} {\bibfnamefont {M.}~\bibnamefont
  {Ashtar}}, \bibinfo {author} {\bibfnamefont {M.~A.}\ \bibnamefont {Marwat}},
  \bibinfo {author} {\bibfnamefont {Y.~X.}\ \bibnamefont {Gao}}, \bibinfo
  {author} {\bibfnamefont {Z.~T.}\ \bibnamefont {Zhang}}, \bibinfo {author}
  {\bibfnamefont {L.}~\bibnamefont {Pi}}, \bibinfo {author} {\bibfnamefont
  {S.~L.}\ \bibnamefont {Yuan}}, \ and\ \bibinfo {author} {\bibfnamefont
  {Z.~M.}\ \bibnamefont {Tian}},\ }\bibfield  {title} {\enquote {\bibinfo
  {title} {REZnAl$_{11}$O$_{19}$ (Re = Pr, Nd, Sm-Tb): a new family of ideal 2D
  triangular lattice frustrated magnets},}\ }\href {\doibase
  10.1039/C9TC02643F} {\bibfield  {journal} {\bibinfo  {journal} {J. Mater.
  Chem. C}\ }\textbf {\bibinfo {volume} {7}},\ \bibinfo {pages} {10073}
  (\bibinfo {year} {2019}{\natexlab{b}})}\BibitemShut {NoStop}%
\bibitem [{\citenamefont {Bu}\ \emph {et~al.}(2022)\citenamefont {Bu},
  \citenamefont {Ashtar}, \citenamefont {Shiroka}, \citenamefont {Walker},
  \citenamefont {Fu}, \citenamefont {Zhao}, \citenamefont {Gardner},
  \citenamefont {Chen}, \citenamefont {Tian},\ and\ \citenamefont
  {Guo}}]{PhysRevB.106.134428}%
  \BibitemOpen
  \bibfield  {author} {\bibinfo {author} {\bibfnamefont {Huanpeng}\
  \bibnamefont {Bu}}, \bibinfo {author} {\bibfnamefont {Malik}\ \bibnamefont
  {Ashtar}}, \bibinfo {author} {\bibfnamefont {Toni}\ \bibnamefont {Shiroka}},
  \bibinfo {author} {\bibfnamefont {Helen~C.}\ \bibnamefont {Walker}}, \bibinfo
  {author} {\bibfnamefont {Zhendong}\ \bibnamefont {Fu}}, \bibinfo {author}
  {\bibfnamefont {Jinkui}\ \bibnamefont {Zhao}}, \bibinfo {author}
  {\bibfnamefont {Jason~S.}\ \bibnamefont {Gardner}}, \bibinfo {author}
  {\bibfnamefont {Gang}\ \bibnamefont {Chen}}, \bibinfo {author} {\bibfnamefont
  {Zhaoming}\ \bibnamefont {Tian}}, \ and\ \bibinfo {author} {\bibfnamefont
  {Hanjie}\ \bibnamefont {Guo}},\ }\bibfield  {title} {\enquote {\bibinfo
  {title} {{Gapless triangular-lattice spin-liquid candidate
  ${\mathrm{PrZnAl}}_{11}{\mathrm{O}}_{19}$}},}\ }\href {\doibase
  10.1103/PhysRevB.106.134428} {\bibfield  {journal} {\bibinfo  {journal}
  {Phys. Rev. B}\ }\textbf {\bibinfo {volume} {106}},\ \bibinfo {pages}
  {134428} (\bibinfo {year} {2022})}\BibitemShut {NoStop}%
\bibitem [{\citenamefont {den Hertog}\ and\ \citenamefont
  {Gingras}(2000)}]{PhysRevLett.84.3430}%
  \BibitemOpen
  \bibfield  {author} {\bibinfo {author} {\bibfnamefont {Byron~C.}\
  \bibnamefont {den Hertog}}\ and\ \bibinfo {author} {\bibfnamefont {Michel
  J.~P.}\ \bibnamefont {Gingras}},\ }\bibfield  {title} {\enquote {\bibinfo
  {title} {Dipolar Interactions and Origin of Spin Ice in Ising Pyrochlore
  Magnets},}\ }\href {\doibase 10.1103/PhysRevLett.84.3430} {\bibfield
  {journal} {\bibinfo  {journal} {Phys. Rev. Lett.}\ }\textbf {\bibinfo
  {volume} {84}},\ \bibinfo {pages} {3430} (\bibinfo {year}
  {2000})}\BibitemShut {NoStop}%
\bibitem [{\citenamefont {Ranjith}\ \emph {et~al.}(2019)\citenamefont
  {Ranjith}, \citenamefont {Dmytriieva}, \citenamefont {Khim}, \citenamefont
  {Sichelschmidt}, \citenamefont {Luther}, \citenamefont {Ehlers},
  \citenamefont {Yasuoka}, \citenamefont {Wosnitza}, \citenamefont {Tsirlin},
  \citenamefont {K\"uhne},\ and\ \citenamefont {Baenitz}}]{PhysRevB.99.180401}%
  \BibitemOpen
  \bibfield  {author} {\bibinfo {author} {\bibfnamefont {K.~M.}\ \bibnamefont
  {Ranjith}}, \bibinfo {author} {\bibfnamefont {D.}~\bibnamefont {Dmytriieva}},
  \bibinfo {author} {\bibfnamefont {S.}~\bibnamefont {Khim}}, \bibinfo {author}
  {\bibfnamefont {J.}~\bibnamefont {Sichelschmidt}}, \bibinfo {author}
  {\bibfnamefont {S.}~\bibnamefont {Luther}}, \bibinfo {author} {\bibfnamefont
  {D.}~\bibnamefont {Ehlers}}, \bibinfo {author} {\bibfnamefont
  {H.}~\bibnamefont {Yasuoka}}, \bibinfo {author} {\bibfnamefont
  {J.}~\bibnamefont {Wosnitza}}, \bibinfo {author} {\bibfnamefont {A.~A.}\
  \bibnamefont {Tsirlin}}, \bibinfo {author} {\bibfnamefont {H.}~\bibnamefont
  {K\"uhne}}, \ and\ \bibinfo {author} {\bibfnamefont {M.}~\bibnamefont
  {Baenitz}},\ }\bibfield  {title} {\enquote {\bibinfo {title} {Field-induced
  instability of the quantum spin liquid ground state in the
  ${J}_{\mathrm{eff}}=\frac{1}{2}$ triangular-lattice compound
  ${\mathrm{NaYbO}}_{2}$},}\ }\href {\doibase 10.1103/PhysRevB.99.180401}
  {\bibfield  {journal} {\bibinfo  {journal} {Phys. Rev. B}\ }\textbf {\bibinfo
  {volume} {99}},\ \bibinfo {pages} {180401(R)} (\bibinfo {year}
  {2019})}\BibitemShut {NoStop}%
\bibitem [{\citenamefont {Yamashita}\ \emph {et~al.}(2008)\citenamefont
  {Yamashita}, \citenamefont {Nakazawa}, \citenamefont {Oguni}, \citenamefont
  {Oshima}, \citenamefont {Nojiri}, \citenamefont {Shimizu}, \citenamefont
  {Miyagawa},\ and\ \citenamefont {Kanoda}}]{np4_459}%
  \BibitemOpen
  \bibfield  {author} {\bibinfo {author} {\bibfnamefont {Satoshi}\ \bibnamefont
  {Yamashita}}, \bibinfo {author} {\bibfnamefont {Yasuhiro}\ \bibnamefont
  {Nakazawa}}, \bibinfo {author} {\bibfnamefont {Masaharu}\ \bibnamefont
  {Oguni}}, \bibinfo {author} {\bibfnamefont {Yugo}\ \bibnamefont {Oshima}},
  \bibinfo {author} {\bibfnamefont {Hiroyuki}\ \bibnamefont {Nojiri}}, \bibinfo
  {author} {\bibfnamefont {Yasuhiro}\ \bibnamefont {Shimizu}}, \bibinfo
  {author} {\bibfnamefont {Kazuya}\ \bibnamefont {Miyagawa}}, \ and\ \bibinfo
  {author} {\bibfnamefont {Kazushi}\ \bibnamefont {Kanoda}},\ }\bibfield
  {title} {\enquote {\bibinfo {title} {{Thermodynamic properties of a spin-1/2
  spin-liquid state in a $\kappa$-type organic salt}},}\ }\href@noop {}
  {\bibfield  {journal} {\bibinfo  {journal} {Nat. Phys.}\ }\textbf {\bibinfo
  {volume} {4}},\ \bibinfo {pages} {459} (\bibinfo {year}
  {2008})}\BibitemShut {NoStop}%
\bibitem [{\citenamefont {Helton}\ \emph {et~al.}(2007)\citenamefont {Helton},
  \citenamefont {Matan}, \citenamefont {Shores}, \citenamefont {Nytko},
  \citenamefont {Bartlett}, \citenamefont {Yoshida}, \citenamefont {Takano},
  \citenamefont {Suslov}, \citenamefont {Qiu}, \citenamefont {Chung},
  \citenamefont {Nocera},\ and\ \citenamefont {Lee}}]{prl98_107204}%
  \BibitemOpen
  \bibfield  {author} {\bibinfo {author} {\bibfnamefont {J.~S.}\ \bibnamefont
  {Helton}}, \bibinfo {author} {\bibfnamefont {K.}~\bibnamefont {Matan}},
  \bibinfo {author} {\bibfnamefont {M.~P.}\ \bibnamefont {Shores}}, \bibinfo
  {author} {\bibfnamefont {E.~A.}\ \bibnamefont {Nytko}}, \bibinfo {author}
  {\bibfnamefont {B.~M.}\ \bibnamefont {Bartlett}}, \bibinfo {author}
  {\bibfnamefont {Y.}~\bibnamefont {Yoshida}}, \bibinfo {author} {\bibfnamefont
  {Y.}~\bibnamefont {Takano}}, \bibinfo {author} {\bibfnamefont
  {A.}~\bibnamefont {Suslov}}, \bibinfo {author} {\bibfnamefont
  {Y.}~\bibnamefont {Qiu}}, \bibinfo {author} {\bibfnamefont {J.-H.}\
  \bibnamefont {Chung}}, \bibinfo {author} {\bibfnamefont {D.~G.}\ \bibnamefont
  {Nocera}}, \ and\ \bibinfo {author} {\bibfnamefont {Y.~S.}\ \bibnamefont
  {Lee}},\ }\bibfield  {title} {\enquote {\bibinfo {title} {{Spin Dynamics of
  the Spin-$1/2$ Kagome Lattice Antiferromagnet
  ${\mathrm{ZnCu}}_{3}(\mathrm{OH}{)}_{6}{\mathrm{Cl}}_{2}$}},}\ }\href@noop {}
  {\bibfield  {journal} {\bibinfo  {journal} {Phys. Rev. Lett.}\ }\textbf
  {\bibinfo {volume} {98}},\ \bibinfo {pages} {107204} (\bibinfo {year}
  {2007})}\BibitemShut {NoStop}%
\bibitem [{\citenamefont {Yamashita}\ \emph {et~al.}(2011)\citenamefont
  {Yamashita}, \citenamefont {Yamamoto}, \citenamefont {Nakazawa},
  \citenamefont {Tamura},\ and\ \citenamefont {Kato}}]{nc2_275}%
  \BibitemOpen
  \bibfield  {author} {\bibinfo {author} {\bibfnamefont {Satoshi}\ \bibnamefont
  {Yamashita}}, \bibinfo {author} {\bibfnamefont {Takashi}\ \bibnamefont
  {Yamamoto}}, \bibinfo {author} {\bibfnamefont {Yasuhiro}\ \bibnamefont
  {Nakazawa}}, \bibinfo {author} {\bibfnamefont {Masafumi}\ \bibnamefont
  {Tamura}}, \ and\ \bibinfo {author} {\bibfnamefont {Reizo}\ \bibnamefont
  {Kato}},\ }\bibfield  {title} {\enquote {\bibinfo {title} {{Gapless spin
  liquid of an organic triangular compound evidenced by thermodynamic
  measurements}},}\ }\href@noop {} {\bibfield  {journal} {\bibinfo  {journal}
  {Nat. Commun.}\ }\textbf {\bibinfo {volume} {2}},\ \bibinfo {pages} {275}
  (\bibinfo {year} {2011})}\BibitemShut {NoStop}%
\bibitem [{\citenamefont {Ran}\ \emph {et~al.}(2007)\citenamefont {Ran},
  \citenamefont {Hermele}, \citenamefont {Lee},\ and\ \citenamefont
  {Wen}}]{PhysRevLett.98.117205}%
  \BibitemOpen
  \bibfield  {author} {\bibinfo {author} {\bibfnamefont {Ying}\ \bibnamefont
  {Ran}}, \bibinfo {author} {\bibfnamefont {Michael}\ \bibnamefont {Hermele}},
  \bibinfo {author} {\bibfnamefont {Patrick~A.}\ \bibnamefont {Lee}}, \ and\
  \bibinfo {author} {\bibfnamefont {Xiao-Gang}\ \bibnamefont {Wen}},\
  }\bibfield  {title} {\enquote {\bibinfo {title} {{Projected-Wave-Function
  Study of the Spin-$1/2$ Heisenberg Model on the Kagom\'e Lattice}},}\ }\href
  {\doibase 10.1103/PhysRevLett.98.117205} {\bibfield  {journal} {\bibinfo
  {journal} {Phys. Rev. Lett.}\ }\textbf {\bibinfo {volume} {98}},\ \bibinfo
  {pages} {117205} (\bibinfo {year} {2007})}\BibitemShut {NoStop}%
\bibitem [{\citenamefont {Hu}\ \emph {et~al.}(2019)\citenamefont {Hu},
  \citenamefont {Zhu}, \citenamefont {Eggert},\ and\ \citenamefont
  {He}}]{PhysRevLett.123.207203}%
  \BibitemOpen
  \bibfield  {author} {\bibinfo {author} {\bibfnamefont {Shijie}\ \bibnamefont
  {Hu}}, \bibinfo {author} {\bibfnamefont {W.}~\bibnamefont {Zhu}}, \bibinfo
  {author} {\bibfnamefont {Sebastian}\ \bibnamefont {Eggert}}, \ and\ \bibinfo
  {author} {\bibfnamefont {Yin-Chen}\ \bibnamefont {He}},\ }\bibfield  {title}
  {\enquote {\bibinfo {title} {{Dirac Spin Liquid on the Spin-$1/2$ Triangular
  Heisenberg Antiferromagnet}},}\ }\href {\doibase
  10.1103/PhysRevLett.123.207203} {\bibfield  {journal} {\bibinfo  {journal}
  {Phys. Rev. Lett.}\ }\textbf {\bibinfo {volume} {123}},\ \bibinfo {pages}
  {207203} (\bibinfo {year} {2019})}\BibitemShut {NoStop}%
\bibitem [{\citenamefont {Do}\ \emph {et~al.}(2020)\citenamefont {Do},
  \citenamefont {Lee}, \citenamefont {Kihara}, \citenamefont {Choi},
  \citenamefont {Yoon}, \citenamefont {Kim}, \citenamefont {Cheong},
  \citenamefont {Chen}, \citenamefont {Chou}, \citenamefont {Nojiri},\ and\
  \citenamefont {Choi}}]{PhysRevLett.124.047204}%
  \BibitemOpen
  \bibfield  {author} {\bibinfo {author} {\bibfnamefont {Seung-Hwan}\
  \bibnamefont {Do}}, \bibinfo {author} {\bibfnamefont {C.~H.}\ \bibnamefont
  {Lee}}, \bibinfo {author} {\bibfnamefont {T.}~\bibnamefont {Kihara}},
  \bibinfo {author} {\bibfnamefont {Y.~S.}\ \bibnamefont {Choi}}, \bibinfo
  {author} {\bibfnamefont {Sungwon}\ \bibnamefont {Yoon}}, \bibinfo {author}
  {\bibfnamefont {Kangwon}\ \bibnamefont {Kim}}, \bibinfo {author}
  {\bibfnamefont {Hyeonsik}\ \bibnamefont {Cheong}}, \bibinfo {author}
  {\bibfnamefont {Wei-Tin}\ \bibnamefont {Chen}}, \bibinfo {author}
  {\bibfnamefont {Fangcheng}\ \bibnamefont {Chou}}, \bibinfo {author}
  {\bibfnamefont {H.}~\bibnamefont {Nojiri}}, \ and\ \bibinfo {author}
  {\bibfnamefont {Kwang-Yong}\ \bibnamefont {Choi}},\ }\bibfield  {title}
  {\enquote {\bibinfo {title} {{Randomly Hopping Majorana Fermions in the
  Diluted Kitaev System
  $\ensuremath{\alpha}$-${\mathrm{Ru}}_{0.8}{\mathrm{Ir}}_{0.2}{\mathrm{Cl}}_{3}$}},}\
  }\href {\doibase 10.1103/PhysRevLett.124.047204} {\bibfield  {journal}
  {\bibinfo  {journal} {Phys. Rev. Lett.}\ }\textbf {\bibinfo {volume} {124}},\
  \bibinfo {pages} {047204} (\bibinfo {year} {2020})}\BibitemShut {NoStop}%
\bibitem [{\citenamefont {Zeng}\ \emph {et~al.}(2022)\citenamefont {Zeng},
  \citenamefont {Ma}, \citenamefont {Wu}, \citenamefont {Li}, \citenamefont
  {Tao}, \citenamefont {Lu}, \citenamefont {Chen}, \citenamefont {Mi},
  \citenamefont {Song}, \citenamefont {Cao}, \citenamefont {Che}, \citenamefont
  {Li}, \citenamefont {Li}, \citenamefont {Luo}, \citenamefont {Meng},\ and\
  \citenamefont {Li}}]{PhysRevB.105.L121109}%
  \BibitemOpen
  \bibfield  {author} {\bibinfo {author} {\bibfnamefont {Zhenyuan}\
  \bibnamefont {Zeng}}, \bibinfo {author} {\bibfnamefont {Xiaoyan}\
  \bibnamefont {Ma}}, \bibinfo {author} {\bibfnamefont {Si}~\bibnamefont {Wu}},
  \bibinfo {author} {\bibfnamefont {Hai-Feng}\ \bibnamefont {Li}}, \bibinfo
  {author} {\bibfnamefont {Zhen}\ \bibnamefont {Tao}}, \bibinfo {author}
  {\bibfnamefont {Xingye}\ \bibnamefont {Lu}}, \bibinfo {author} {\bibfnamefont
  {Xiao-hui}\ \bibnamefont {Chen}}, \bibinfo {author} {\bibfnamefont
  {Jin-Xiao}\ \bibnamefont {Mi}}, \bibinfo {author} {\bibfnamefont {Shi-Jie}\
  \bibnamefont {Song}}, \bibinfo {author} {\bibfnamefont {Guang-Han}\
  \bibnamefont {Cao}}, \bibinfo {author} {\bibfnamefont {Guangwei}\
  \bibnamefont {Che}}, \bibinfo {author} {\bibfnamefont {Kuo}\ \bibnamefont
  {Li}}, \bibinfo {author} {\bibfnamefont {Gang}\ \bibnamefont {Li}}, \bibinfo
  {author} {\bibfnamefont {Huiqian}\ \bibnamefont {Luo}}, \bibinfo {author}
  {\bibfnamefont {Zi~Yang}\ \bibnamefont {Meng}}, \ and\ \bibinfo {author}
  {\bibfnamefont {Shiliang}\ \bibnamefont {Li}},\ }\bibfield  {title} {\enquote
  {\bibinfo {title} {{Possible Dirac quantum spin liquid in the kagome quantum
  antiferromagnet
  ${\mathrm{YCu}}_{3}{(\mathrm{OH})}_{6}{\mathrm{Br}}_{2}[{\mathrm{Br}}_{x}{(\mathrm{OH})}_{1\ensuremath{-}x}]$}},}\
  }\href {\doibase 10.1103/PhysRevB.105.L121109} {\bibfield  {journal}
  {\bibinfo  {journal} {Phys. Rev. B}\ }\textbf {\bibinfo {volume} {105}},\
  \bibinfo {pages} {L121109} (\bibinfo {year} {2022})}\BibitemShut {NoStop}%
\bibitem [{\citenamefont {Liu}\ \emph {et~al.}(2022{\natexlab{b}})\citenamefont
  {Liu}, \citenamefont {Zeng}, \citenamefont {Xu}, \citenamefont {Sun},
  \citenamefont {Yakubovich}, \citenamefont {Shvanskaya}, \citenamefont {Li},\
  and\ \citenamefont {Vasiliev}}]{PhysRevB.105.155153}%
  \BibitemOpen
  \bibfield  {author} {\bibinfo {author} {\bibfnamefont {Bo}~\bibnamefont
  {Liu}}, \bibinfo {author} {\bibfnamefont {Zhenyuan}\ \bibnamefont {Zeng}},
  \bibinfo {author} {\bibfnamefont {Aini}\ \bibnamefont {Xu}}, \bibinfo
  {author} {\bibfnamefont {Yili}\ \bibnamefont {Sun}}, \bibinfo {author}
  {\bibfnamefont {Olga}\ \bibnamefont {Yakubovich}}, \bibinfo {author}
  {\bibfnamefont {Larisa}\ \bibnamefont {Shvanskaya}}, \bibinfo {author}
  {\bibfnamefont {Shiliang}\ \bibnamefont {Li}}, \ and\ \bibinfo {author}
  {\bibfnamefont {Alexander}\ \bibnamefont {Vasiliev}},\ }\bibfield  {title}
  {\enquote {\bibinfo {title} {Low-temperature specific-heat studies on two
  square-kagome antiferromagnets},}\ }\href {\doibase
  10.1103/PhysRevB.105.155153} {\bibfield  {journal} {\bibinfo  {journal}
  {Phys. Rev. B}\ }\textbf {\bibinfo {volume} {105}},\ \bibinfo {pages}
  {155153} (\bibinfo {year} {2022}{\natexlab{b}})}\BibitemShut {NoStop}%
\bibitem [{\citenamefont {Xu}\ \emph {et~al.}()\citenamefont {Xu},
  \citenamefont {Bag}, \citenamefont {Sherman}, \citenamefont {Yadav},
  \citenamefont {Kolesnikov}, \citenamefont {Podlesnyak}, \citenamefont
  {Moore},\ and\ \citenamefont {Haravifard}}]{arXiv_2305.20040}%
  \BibitemOpen
  \bibfield  {author} {\bibinfo {author} {\bibfnamefont {Sijie}\ \bibnamefont
  {Xu}}, \bibinfo {author} {\bibfnamefont {Rabindranath}\ \bibnamefont {Bag}},
  \bibinfo {author} {\bibfnamefont {Nicholas~E.}\ \bibnamefont {Sherman}},
  \bibinfo {author} {\bibfnamefont {Lalit}\ \bibnamefont {Yadav}}, \bibinfo
  {author} {\bibfnamefont {Alexander~I.}\ \bibnamefont {Kolesnikov}}, \bibinfo
  {author} {\bibfnamefont {Andrey~A.}\ \bibnamefont {Podlesnyak}}, \bibinfo
  {author} {\bibfnamefont {Joel~E.}\ \bibnamefont {Moore}}, \ and\ \bibinfo
  {author} {\bibfnamefont {Sara}\ \bibnamefont {Haravifard}},\ }\bibfield
  {title} {\enquote {\bibinfo {title} {{Realization of U(1) Dirac Quantum Spin
  Liquid in YbZn$_2$GaO$_5$}},}\ }\href@noop {} {\bibinfo  {journal}
  {arXiv:2305.20040}\ }\BibitemShut {NoStop}%
\bibitem [{\citenamefont {Clark}\ \emph {et~al.}(2019)\citenamefont {Clark},
  \citenamefont {Sala}, \citenamefont {Maharaj}, \citenamefont {Stone},
  \citenamefont {Knight}, \citenamefont {Telling}, \citenamefont {Wang},
  \citenamefont {Xu}, \citenamefont {Kim}, \citenamefont {Li}, \citenamefont
  {Cheong},\ and\ \citenamefont {Gaulin}}]{np15_262}%
  \BibitemOpen
\bibfield  {journal} {  }\bibfield  {author} {\bibinfo {author} {\bibfnamefont
  {Lucy}\ \bibnamefont {Clark}}, \bibinfo {author} {\bibfnamefont {Gabriele}\
  \bibnamefont {Sala}}, \bibinfo {author} {\bibfnamefont {Dalini~D.}\
  \bibnamefont {Maharaj}}, \bibinfo {author} {\bibfnamefont {Matthew~B.}\
  \bibnamefont {Stone}}, \bibinfo {author} {\bibfnamefont {Kevin~S.}\
  \bibnamefont {Knight}}, \bibinfo {author} {\bibfnamefont {Mark T.~F.}\
  \bibnamefont {Telling}}, \bibinfo {author} {\bibfnamefont {Xueyun}\
  \bibnamefont {Wang}}, \bibinfo {author} {\bibfnamefont {Xianghan}\
  \bibnamefont {Xu}}, \bibinfo {author} {\bibfnamefont {Jaewook}\ \bibnamefont
  {Kim}}, \bibinfo {author} {\bibfnamefont {Yanbin}\ \bibnamefont {Li}},
  \bibinfo {author} {\bibfnamefont {Sang-Wook}\ \bibnamefont {Cheong}}, \ and\
  \bibinfo {author} {\bibfnamefont {Bruce~D.}\ \bibnamefont {Gaulin}},\
  }\bibfield  {title} {\enquote {\bibinfo {title} {Two-dimensional spin liquid
  behaviour in the triangular-honeycomb antiferromagnet TbInO$_3$},}\ }\href@noop
  {} {\bibfield  {journal} {\bibinfo  {journal} {Nat. Phys.}\ }\textbf
  {\bibinfo {volume} {15}},\ \bibinfo {pages} {262} (\bibinfo {year}
  {2019})}\BibitemShut {NoStop}%
\bibitem [{\citenamefont {Scheie}\ \emph {et~al.}(2018)\citenamefont {Scheie},
  \citenamefont {Sanders}, \citenamefont {Krizan}, \citenamefont
  {Christianson}, \citenamefont {Garlea}, \citenamefont {Cava},\ and\
  \citenamefont {Broholm}}]{PhysRevB.98.134401}%
  \BibitemOpen
  \bibfield  {author} {\bibinfo {author} {\bibfnamefont {A.}~\bibnamefont
  {Scheie}}, \bibinfo {author} {\bibfnamefont {M.}~\bibnamefont {Sanders}},
  \bibinfo {author} {\bibfnamefont {J.}~\bibnamefont {Krizan}}, \bibinfo
  {author} {\bibfnamefont {A.~D.}\ \bibnamefont {Christianson}}, \bibinfo
  {author} {\bibfnamefont {V.~O.}\ \bibnamefont {Garlea}}, \bibinfo {author}
  {\bibfnamefont {R.~J.}\ \bibnamefont {Cava}}, \ and\ \bibinfo {author}
  {\bibfnamefont {C.}~\bibnamefont {Broholm}},\ }\bibfield  {title} {\enquote
  {\bibinfo {title} {{Crystal field levels and magnetic anisotropy in the
  kagome compounds
  ${\mathrm{Nd}}_{3}{\mathrm{Sb}}_{3}{\mathrm{Mg}}_{2}{\mathrm{O}}_{14}$,
  ${\mathrm{Nd}}_{3}{\mathrm{Sb}}_{3}{\mathrm{Zn}}_{2}{\mathrm{O}}_{14}$, and
  ${\mathrm{Pr}}_{3}{\mathrm{Sb}}_{3}{\mathrm{Mg}}_{2}{\mathrm{O}}_{14}$}},}\
  }\href {\doibase 10.1103/PhysRevB.98.134401} {\bibfield  {journal} {\bibinfo
  {journal} {Phys. Rev. B}\ }\textbf {\bibinfo {volume} {98}},\ \bibinfo
  {pages} {134401} (\bibinfo {year} {2018})}\BibitemShut {NoStop}%
\bibitem [{\citenamefont {Gaudet}\ \emph {et~al.}(2019)\citenamefont {Gaudet},
  \citenamefont {Smith}, \citenamefont {Dudemaine}, \citenamefont {Beare},
  \citenamefont {Buhariwalla}, \citenamefont {Butch}, \citenamefont {Stone},
  \citenamefont {Kolesnikov}, \citenamefont {Xu}, \citenamefont {Yahne},
  \citenamefont {Ross}, \citenamefont {Marjerrison}, \citenamefont {Garrett},
  \citenamefont {Luke}, \citenamefont {Bianchi},\ and\ \citenamefont
  {Gaulin}}]{PhysRevLett.122.187201}%
  \BibitemOpen
  \bibfield  {author} {\bibinfo {author} {\bibfnamefont {J.}~\bibnamefont
  {Gaudet}}, \bibinfo {author} {\bibfnamefont {E.~M.}\ \bibnamefont {Smith}},
  \bibinfo {author} {\bibfnamefont {J.}~\bibnamefont {Dudemaine}}, \bibinfo
  {author} {\bibfnamefont {J.}~\bibnamefont {Beare}}, \bibinfo {author}
  {\bibfnamefont {C.~R.~C.}\ \bibnamefont {Buhariwalla}}, \bibinfo {author}
  {\bibfnamefont {N.~P.}\ \bibnamefont {Butch}}, \bibinfo {author}
  {\bibfnamefont {M.~B.}\ \bibnamefont {Stone}}, \bibinfo {author}
  {\bibfnamefont {A.~I.}\ \bibnamefont {Kolesnikov}}, \bibinfo {author}
  {\bibfnamefont {Guangyong}\ \bibnamefont {Xu}}, \bibinfo {author}
  {\bibfnamefont {D.~R.}\ \bibnamefont {Yahne}}, \bibinfo {author}
  {\bibfnamefont {K.~A.}\ \bibnamefont {Ross}}, \bibinfo {author}
  {\bibfnamefont {C.~A.}\ \bibnamefont {Marjerrison}}, \bibinfo {author}
  {\bibfnamefont {J.~D.}\ \bibnamefont {Garrett}}, \bibinfo {author}
  {\bibfnamefont {G.~M.}\ \bibnamefont {Luke}}, \bibinfo {author}
  {\bibfnamefont {A.~D.}\ \bibnamefont {Bianchi}}, \ and\ \bibinfo {author}
  {\bibfnamefont {B.~D.}\ \bibnamefont {Gaulin}},\ }\bibfield  {title}
  {\enquote {\bibinfo {title} {{Quantum Spin Ice Dynamics in the
  Dipole-Octupole Pyrochlore Magnet
  ${\mathrm{Ce}}_{2}{\mathrm{Zr}}_{2}{\mathrm{O}}_{7}$}},}\ }\href {\doibase
  10.1103/PhysRevLett.122.187201} {\bibfield  {journal} {\bibinfo  {journal}
  {Phys. Rev. Lett.}\ }\textbf {\bibinfo {volume} {122}},\ \bibinfo {pages}
  {187201} (\bibinfo {year} {2019})}\BibitemShut {NoStop}%
\bibitem [{\citenamefont {Ortiz}\ \emph {et~al.}(2023)\citenamefont {Ortiz},
  \citenamefont {Sarte}, \citenamefont {Avidor}, \citenamefont {Hay},
  \citenamefont {Kenney}, \citenamefont {Kolesnikov}, \citenamefont
  {Pajerowski}, \citenamefont {Aczel}, \citenamefont {Taddei}, \citenamefont
  {Brown}, \citenamefont {Wang}, \citenamefont {Graf}, \citenamefont
  {Seshadri}, \citenamefont {Balents},\ and\ \citenamefont
  {Wilson}}]{np19_949}%
  \BibitemOpen
  \bibfield  {author} {\bibinfo {author} {\bibfnamefont {Brenden~R.}\
  \bibnamefont {Ortiz}}, \bibinfo {author} {\bibfnamefont {Paul~M.}\
  \bibnamefont {Sarte}}, \bibinfo {author} {\bibfnamefont {Alon~Hendler}\
  \bibnamefont {Avidor}}, \bibinfo {author} {\bibfnamefont {Aurland}\
  \bibnamefont {Hay}}, \bibinfo {author} {\bibfnamefont {Eric}\ \bibnamefont
  {Kenney}}, \bibinfo {author} {\bibfnamefont {Alexander~I.}\ \bibnamefont
  {Kolesnikov}}, \bibinfo {author} {\bibfnamefont {Daniel~M.}\ \bibnamefont
  {Pajerowski}}, \bibinfo {author} {\bibfnamefont {Adam~A.}\ \bibnamefont
  {Aczel}}, \bibinfo {author} {\bibfnamefont {Keith~M.}\ \bibnamefont
  {Taddei}}, \bibinfo {author} {\bibfnamefont {Craig~M.}\ \bibnamefont
  {Brown}}, \bibinfo {author} {\bibfnamefont {Chennan}\ \bibnamefont {Wang}},
  \bibinfo {author} {\bibfnamefont {Michael~J.}\ \bibnamefont {Graf}}, \bibinfo
  {author} {\bibfnamefont {Ram}\ \bibnamefont {Seshadri}}, \bibinfo {author}
  {\bibfnamefont {Leon}\ \bibnamefont {Balents}}, \ and\ \bibinfo {author}
  {\bibfnamefont {Stephen~D.}\ \bibnamefont {Wilson}},\ }\bibfield  {title}
  {\enquote {\bibinfo {title} {{Quantum disordered ground state in the
  triangular-lattice magnet NaRuO$_2$}},}\ }\href@noop {} {\bibfield  {journal}
  {\bibinfo  {journal} {Nat. Phys.}\ }\textbf {\bibinfo {volume} {19}},\
  \bibinfo {pages} {943} (\bibinfo {year} {2023})}\BibitemShut {NoStop}%
\bibitem [{\citenamefont {Hutchings}(1964)}]{HUTCHINGS1964227}%
  \BibitemOpen
  \bibfield  {author} {\bibinfo {author} {\bibfnamefont {M.T.}\ \bibnamefont
  {Hutchings}},\ }\bibfield  {title} {\enquote {\bibinfo {title} {Point-Charge
  Calculations of Energy levels of Magnetic Ions in Crystalline Electric
  Fields},}\ \
  }(\bibinfo  {publisher} {Academic Press, New York},\ \bibinfo {year} {1964})\ pp.\
  \bibinfo {pages} {227--273}\BibitemShut {NoStop}%
\bibitem [{\citenamefont {Stevens}(1952)}]{Stevens_1952}%
  \BibitemOpen
  \bibfield  {author} {\bibinfo {author} {\bibfnamefont {K.~W.~H.}\ \bibnamefont
  {Stevens}},\ }\bibfield  {title} {\enquote {\bibinfo {title} {Matrix Elements
  and Operator Equivalents Connected with the Magnetic Properties of Rare Earth
  Ions},}\ }\href {\doibase 10.1088/0370-1298/65/3/308} {\bibfield  {journal}
  {\bibinfo  {journal} {Proc. Phys. Soc. Sec. A}\
  }\textbf {\bibinfo {volume} {65}},\ \bibinfo {pages} {209} (\bibinfo
  {year} {1952})}\BibitemShut {NoStop}%
\bibitem [{\citenamefont {Edvardsson}\ and\ \citenamefont
  {Klintenberg}(1998)}]{EDVARDSSON1998230}%
  \BibitemOpen
  \bibfield  {author} {\bibinfo {author} {\bibfnamefont {Sverker}\ \bibnamefont
  {Edvardsson}}\ and\ \bibinfo {author} {\bibfnamefont {Mattias}\ \bibnamefont
  {Klintenberg}},\ }\bibfield  {title} {\enquote {\bibinfo {title} {{Role of
  the electrostatic model in calculating rare-earth crystal-field
  parameters}},}\ }\href {\doibase
  https://doi.org/10.1016/S0925-8388(98)00309-0} {\bibfield  {journal}
  {\bibinfo  {journal} {J. Alloy. Compd.}\ }\textbf {\bibinfo
  {volume} {275-277}},\ \bibinfo {pages} {230} (\bibinfo {year}
  {1998})}\BibitemShut {NoStop}%
\bibitem [{\citenamefont {Liu}\ \emph {et~al.}(2018{\natexlab{b}})\citenamefont
  {Liu}, \citenamefont {Li},\ and\ \citenamefont {Chen}}]{PhysRevB.98.045119}%
  \BibitemOpen
  \bibfield  {author} {\bibinfo {author} {\bibfnamefont {Changle}\ \bibnamefont
  {Liu}}, \bibinfo {author} {\bibfnamefont {Yao-Dong}\ \bibnamefont {Li}}, \
  and\ \bibinfo {author} {\bibfnamefont {Gang}\ \bibnamefont {Chen}},\
  }\bibfield  {title} {\enquote {\bibinfo {title} {{Selective measurements of
  intertwined multipolar orders: Non-Kramers doublets on a triangular
  lattice}},}\ }\href {\doibase 10.1103/PhysRevB.98.045119} {\bibfield
  {journal} {\bibinfo  {journal} {Phys. Rev. B}\ }\textbf {\bibinfo {volume}
  {98}},\ \bibinfo {pages} {045119} (\bibinfo {year}
  {2018}{\natexlab{b}})}\BibitemShut {NoStop}%
\bibitem [{\citenamefont {Dun}\ \emph {et~al.}(2021)\citenamefont {Dun},
  \citenamefont {Bai}, \citenamefont {Stone}, \citenamefont {Zhou},\ and\
  \citenamefont {Mourigal}}]{PhysRevResearch.3.023012}%
  \BibitemOpen
  \bibfield  {author} {\bibinfo {author} {\bibfnamefont {Zhiling}\ \bibnamefont
  {Dun}}, \bibinfo {author} {\bibfnamefont {Xiaojian}\ \bibnamefont {Bai}},
  \bibinfo {author} {\bibfnamefont {Matthew~B.}\ \bibnamefont {Stone}},
  \bibinfo {author} {\bibfnamefont {Haidong}\ \bibnamefont {Zhou}}, \ and\
  \bibinfo {author} {\bibfnamefont {Martin}\ \bibnamefont {Mourigal}},\
  }\bibfield  {title} {\enquote {\bibinfo {title} {{Effective point-charge
  analysis of crystal fields: Application to rare-earth pyrochlores and tripod
  kagome magnets
  $R{}_{3}\mathrm{Mg}{}_{2}\mathrm{Sb}{}_{3}\mathrm{O}{}_{14}$}},}\ }\href
  {\doibase 10.1103/PhysRevResearch.3.023012} {\bibfield  {journal} {\bibinfo
  {journal} {Phys. Rev. Res.}\ }\textbf {\bibinfo {volume} {3}},\ \bibinfo
  {pages} {023012} (\bibinfo {year} {2021})}\BibitemShut {NoStop}%
\bibitem [{\citenamefont {Li}\ \emph {et~al.}(2018)\citenamefont {Li},
  \citenamefont {Bachus}, \citenamefont {Tokiwa}, \citenamefont {Tsirlin},\
  and\ \citenamefont {Gegenwart}}]{PhysRevB.97.184434}%
  \BibitemOpen
  \bibfield  {author} {\bibinfo {author} {\bibfnamefont {Yuesheng}\
  \bibnamefont {Li}}, \bibinfo {author} {\bibfnamefont {Sebastian}\
  \bibnamefont {Bachus}}, \bibinfo {author} {\bibfnamefont {Yoshifumi}\
  \bibnamefont {Tokiwa}}, \bibinfo {author} {\bibfnamefont {Alexander~A.}\
  \bibnamefont {Tsirlin}}, \ and\ \bibinfo {author} {\bibfnamefont {Philipp}\
  \bibnamefont {Gegenwart}},\ }\bibfield  {title} {\enquote {\bibinfo {title}
  {Gapped ground state in the zigzag pseudospin-1/2 quantum antiferromagnetic
  chain compound ${\mathrm{PrTiNbO}}_{6}$},}\ }\href {\doibase
  10.1103/PhysRevB.97.184434} {\bibfield  {journal} {\bibinfo  {journal} {Phys.
  Rev. B}\ }\textbf {\bibinfo {volume} {97}},\ \bibinfo {pages} {184434}
  (\bibinfo {year} {2018})}\BibitemShut {NoStop}%
\bibitem [{\citenamefont {Nagl}\ \emph {et~al.}(2024)\citenamefont {Nagl},
  \citenamefont {Flavi\'{a}n}, \citenamefont {Hayashida}, \citenamefont {Povarov},
  \citenamefont {Yan}, \citenamefont {Murai}, \citenamefont {Ohira-Kawamura},
  \citenamefont {Simutis}, \citenamefont {Hicken}, \citenamefont {Luetkens},
  \citenamefont {Baines}, \citenamefont {Hauspurg}, \citenamefont {Schwarze},
  \citenamefont {Husstedt}, \citenamefont {Pomjakushin}, \citenamefont
  {Fennell}, \citenamefont {Yan}, \citenamefont {Gvasaliya},\ and\
  \citenamefont {Zheludev}}]{nagl2024excitation}%
  \BibitemOpen
  \bibfield  {author} {\bibinfo {author} {\bibfnamefont {J.}~\bibnamefont
  {Nagl}}, \bibinfo {author} {\bibfnamefont {D.}~\bibnamefont {Flavi\'{a}n}},
  \bibinfo {author} {\bibfnamefont {S.}~\bibnamefont {Hayashida}}, \bibinfo
  {author} {\bibfnamefont {K.~Yu.}\ \bibnamefont {Povarov}}, \bibinfo {author}
  {\bibfnamefont {M.}~\bibnamefont {Yan}}, \bibinfo {author} {\bibfnamefont
  {N.}~\bibnamefont {Murai}}, \bibinfo {author} {\bibfnamefont
  {S.}~\bibnamefont {Ohira-Kawamura}}, \bibinfo {author} {\bibfnamefont
  {G.}~\bibnamefont {Simutis}}, \bibinfo {author} {\bibfnamefont {T.~J.}\
  \bibnamefont {Hicken}}, \bibinfo {author} {\bibfnamefont {H.}~\bibnamefont
  {Luetkens}}, \bibinfo {author} {\bibfnamefont {C.}~\bibnamefont {Baines}},
  \bibinfo {author} {\bibfnamefont {A.}~\bibnamefont {Hauspurg}}, \bibinfo
  {author} {\bibfnamefont {B.~V.}\ \bibnamefont {Schwarze}}, \bibinfo {author}
  {\bibfnamefont {F.}~\bibnamefont {Husstedt}}, \bibinfo {author}
  {\bibfnamefont {V.}~\bibnamefont {Pomjakushin}}, \bibinfo {author}
  {\bibfnamefont {T.}~\bibnamefont {Fennell}}, \bibinfo {author} {\bibfnamefont
  {Z.}~\bibnamefont {Yan}}, \bibinfo {author} {\bibfnamefont {S.}~\bibnamefont
  {Gvasaliya}}, \ and\ \bibinfo {author} {\bibfnamefont {A.}~\bibnamefont
  {Zheludev}},\ }\bibfield  {title} {\enquote {\bibinfo {title} {{Excitation
  Spectrum and Spin Hamiltonian of the Frustrated Quantum Ising Magnet
  Pr$_3$BWO$_9$}},}\ }\href@noop {} {\bibinfo {journal} {arXiv:2402.14107}\ }\BibitemShut {NoStop}%
\bibitem [{\citenamefont {Huang}\ \emph {et~al.}(2021)\citenamefont {Huang},
  \citenamefont {Xu}, \citenamefont {Wang}, \citenamefont {Zhao}, \citenamefont
  {Tu}, \citenamefont {Ni}, \citenamefont {Wang}, \citenamefont {Pan},
  \citenamefont {Fu}, \citenamefont {Hao}, \citenamefont {Liu}, \citenamefont
  {Mei},\ and\ \citenamefont {Li}}]{PhysRevLett.127.267202}%
  \BibitemOpen
  \bibfield  {author} {\bibinfo {author} {\bibfnamefont {Y.~Y.}\ \bibnamefont
  {Huang}}, \bibinfo {author} {\bibfnamefont {Y.}~\bibnamefont {Xu}}, \bibinfo
  {author} {\bibfnamefont {Le}~\bibnamefont {Wang}}, \bibinfo {author}
  {\bibfnamefont {C.~C.}\ \bibnamefont {Zhao}}, \bibinfo {author}
  {\bibfnamefont {C.~P.}\ \bibnamefont {Tu}}, \bibinfo {author} {\bibfnamefont
  {J.~M.}\ \bibnamefont {Ni}}, \bibinfo {author} {\bibfnamefont {L.~S.}\
  \bibnamefont {Wang}}, \bibinfo {author} {\bibfnamefont {B.~L.}\ \bibnamefont
  {Pan}}, \bibinfo {author} {\bibfnamefont {Ying}\ \bibnamefont {Fu}}, \bibinfo
  {author} {\bibfnamefont {Zhanyang}\ \bibnamefont {Hao}}, \bibinfo {author}
  {\bibfnamefont {Cai}\ \bibnamefont {Liu}}, \bibinfo {author} {\bibfnamefont
  {Jia-Wei}\ \bibnamefont {Mei}}, \ and\ \bibinfo {author} {\bibfnamefont
  {S.~Y.}\ \bibnamefont {Li}},\ }\bibfield  {title} {\enquote {\bibinfo {title}
  {{Heat Transport in Herbertsmithite: Can a Quantum Spin Liquid Survive
  Disorder?}}}\ }\href {\doibase 10.1103/PhysRevLett.127.267202} {\bibfield
  {journal} {\bibinfo  {journal} {Phys. Rev. Lett.}\ }\textbf {\bibinfo
  {volume} {127}},\ \bibinfo {pages} {267202} (\bibinfo {year}
  {2021})}\BibitemShut {NoStop}%
\bibitem [{\citenamefont {Norman}(2016)}]{RevModPhys.88.041002}%
  \BibitemOpen
  \bibfield  {author} {\bibinfo {author} {\bibfnamefont {M.~R.}\ \bibnamefont
  {Norman}},\ }\bibfield  {title} {\enquote {\bibinfo {title}
  {\textit{Colloquium} : Herbertsmithite and the search for the quantum spin
  liquid},}\ }\href {\doibase 10.1103/RevModPhys.88.041002} {\bibfield
  {journal} {\bibinfo  {journal} {Rev. Mod. Phys.}\ }\textbf {\bibinfo {volume}
  {88}},\ \bibinfo {pages} {041002} (\bibinfo {year} {2016})}\BibitemShut
  {NoStop}%
\bibitem [{\citenamefont {Li}\ \emph {et~al.}(2020)\citenamefont {Li},
  \citenamefont {Bachus}, \citenamefont {Deng}, \citenamefont {Schmidt},
  \citenamefont {Thoma}, \citenamefont {Hutanu}, \citenamefont {Tokiwa},
  \citenamefont {Tsirlin},\ and\ \citenamefont
  {Gegenwart}}]{PhysRevX.10.011007}%
  \BibitemOpen
  \bibfield  {author} {\bibinfo {author} {\bibfnamefont {Yuesheng}\
  \bibnamefont {Li}}, \bibinfo {author} {\bibfnamefont {Sebastian}\
  \bibnamefont {Bachus}}, \bibinfo {author} {\bibfnamefont {Hao}\ \bibnamefont
  {Deng}}, \bibinfo {author} {\bibfnamefont {Wolfgang}\ \bibnamefont
  {Schmidt}}, \bibinfo {author} {\bibfnamefont {Henrik}\ \bibnamefont {Thoma}},
  \bibinfo {author} {\bibfnamefont {Vladimir}\ \bibnamefont {Hutanu}}, \bibinfo
  {author} {\bibfnamefont {Yoshifumi}\ \bibnamefont {Tokiwa}}, \bibinfo
  {author} {\bibfnamefont {Alexander~A.}\ \bibnamefont {Tsirlin}}, \ and\
  \bibinfo {author} {\bibfnamefont {Philipp}\ \bibnamefont {Gegenwart}},\
  }\bibfield  {title} {\enquote {\bibinfo {title} {Partial Up-Up-Down Order
  with the Continuously Distributed Order Parameter in the Triangular
  Antiferromagnet ${\mathrm{TmMgGaO}}_{4}$},}\ }\href {\doibase
  10.1103/PhysRevX.10.011007} {\bibfield  {journal} {\bibinfo  {journal} {Phys.
  Rev. X}\ }\textbf {\bibinfo {volume} {10}},\ \bibinfo {pages} {011007}
  (\bibinfo {year} {2020})}\BibitemShut {NoStop}%
\bibitem [{\citenamefont {Arh}\ \emph {et~al.}(2022)\citenamefont {Arh},
  \citenamefont {Sana}, \citenamefont {Pregelj}, \citenamefont {Khuntia},
  \citenamefont {Jagli\v{c}i\'{c}}, \citenamefont {Le}, \citenamefont {Biswas},
  \citenamefont {Manuel}, \citenamefont {Mangin-Thro}, \citenamefont
  {Ozarowski},\ and\ \citenamefont {Zorko}}]{nm21_416}%
  \BibitemOpen
  \bibfield  {author} {\bibinfo {author} {\bibfnamefont {T.}~\bibnamefont
  {Arh}}, \bibinfo {author} {\bibfnamefont {B.}~\bibnamefont {Sana}}, \bibinfo
  {author} {\bibfnamefont {M.}~\bibnamefont {Pregelj}}, \bibinfo {author}
  {\bibfnamefont {P.}~\bibnamefont {Khuntia}}, \bibinfo {author} {\bibfnamefont
  {Z.}~\bibnamefont {Jagli\v{c}i\'{c}}}, \bibinfo {author} {\bibfnamefont
  {M.~D.}\ \bibnamefont {Le}}, \bibinfo {author} {\bibfnamefont {P.~K.}\
  \bibnamefont {Biswas}}, \bibinfo {author} {\bibfnamefont {P.}~\bibnamefont
  {Manuel}}, \bibinfo {author} {\bibfnamefont {L.}~\bibnamefont {Mangin-Thro}},
  \bibinfo {author} {\bibfnamefont {A.}~\bibnamefont {Ozarowski}}, \ and\
  \bibinfo {author} {\bibfnamefont {A.}~\bibnamefont {Zorko}},\ }\bibfield
  {title} {\enquote {\bibinfo {title} {{The Ising triangular-lattice
  antiferromagnet neodymium heptatantalate as a quantum spin liquid
  candidate}},}\ }\href@noop {} {\bibfield  {journal} {\bibinfo  {journal}
  {Nat. Mater.}\ }\textbf {\bibinfo {volume} {21}},\ \bibinfo {pages}
  {416} (\bibinfo {year} {2022})}\BibitemShut {NoStop}%
\bibitem [{\citenamefont {Cai}\ \emph {et~al.}(2020)\citenamefont {Cai},
  \citenamefont {Lygouras}, \citenamefont {Thomas}, \citenamefont {Wilson},
  \citenamefont {Beare}, \citenamefont {Sharma}, \citenamefont {Marjerrison},
  \citenamefont {Yahne}, \citenamefont {Ross}, \citenamefont {Gong},
  \citenamefont {Uemura}, \citenamefont {Dabkowska},\ and\ \citenamefont
  {Luke}}]{PhysRevB.101.094432}%
  \BibitemOpen
  \bibfield  {author} {\bibinfo {author} {\bibfnamefont {Y.}~\bibnamefont
  {Cai}}, \bibinfo {author} {\bibfnamefont {C.}~\bibnamefont {Lygouras}},
  \bibinfo {author} {\bibfnamefont {G.}~\bibnamefont {Thomas}}, \bibinfo
  {author} {\bibfnamefont {M.~N.}\ \bibnamefont {Wilson}}, \bibinfo {author}
  {\bibfnamefont {J.}~\bibnamefont {Beare}}, \bibinfo {author} {\bibfnamefont
  {S.}~\bibnamefont {Sharma}}, \bibinfo {author} {\bibfnamefont {C.~A.}\
  \bibnamefont {Marjerrison}}, \bibinfo {author} {\bibfnamefont {D.~R.}\
  \bibnamefont {Yahne}}, \bibinfo {author} {\bibfnamefont {K.~A.}\ \bibnamefont
  {Ross}}, \bibinfo {author} {\bibfnamefont {Z.}~\bibnamefont {Gong}}, \bibinfo
  {author} {\bibfnamefont {Y.~J.}\ \bibnamefont {Uemura}}, \bibinfo {author}
  {\bibfnamefont {H.~A.}\ \bibnamefont {Dabkowska}}, \ and\ \bibinfo {author}
  {\bibfnamefont {G.~M.}\ \bibnamefont {Luke}},\ }\bibfield  {title} {\enquote
  {\bibinfo {title} {{$\ensuremath{\mu}\mathrm{SR}$ study of the triangular
  Ising antiferromagnet ${\mathrm{ErMgGaO}}_{4}$}},}\ }\href {\doibase
  10.1103/PhysRevB.101.094432} {\bibfield  {journal} {\bibinfo  {journal}
  {Phys. Rev. B}\ }\textbf {\bibinfo {volume} {101}},\ \bibinfo {pages}
  {094432} (\bibinfo {year} {2020})}\BibitemShut {NoStop}%
\bibitem [{\citenamefont {Cevallos}\ \emph {et~al.}(2018)\citenamefont
  {Cevallos}, \citenamefont {Stolze},\ and\ \citenamefont
  {Cava}}]{CEVALLOS20185}%
  \BibitemOpen
  \bibfield  {author} {\bibinfo {author} {\bibfnamefont {F.~Alex}\ \bibnamefont
  {Cevallos}}, \bibinfo {author} {\bibfnamefont {Karoline}\ \bibnamefont
  {Stolze}}, \ and\ \bibinfo {author} {\bibfnamefont {Robert~J.}\ \bibnamefont
  {Cava}},\ }\bibfield  {title} {\enquote {\bibinfo {title} {{Structural
  disorder and elementary magnetic properties of triangular lattice ErMgGaO$_4$
  single crystals}},}\ }\href {\doibase
  https://doi.org/10.1016/j.ssc.2018.03.015} {\bibfield  {journal} {\bibinfo
  {journal} {Solid State Commun.}\ }\textbf {\bibinfo {volume} {276}},\
  \bibinfo {pages} {5} (\bibinfo {year} {2018})}\BibitemShut {NoStop}%
\bibitem [{\citenamefont {Ulaga}\ \emph {et~al.}()\citenamefont {Ulaga},
  \citenamefont {Kokalj}, \citenamefont {Wietek}, \citenamefont {Zorko},\ and\
  \citenamefont {Prelovsek}}]{arXiv_2307.03545}%
  \BibitemOpen
  \bibfield  {author} {\bibinfo {author} {\bibfnamefont {M.}~\bibnamefont
  {Ulaga}}, \bibinfo {author} {\bibfnamefont {J.}~\bibnamefont {Kokalj}},
  \bibinfo {author} {\bibfnamefont {A.}~\bibnamefont {Wietek}}, \bibinfo
  {author} {\bibfnamefont {A.}~\bibnamefont {Zorko}}, \ and\ \bibinfo {author}
  {\bibfnamefont {P.}~\bibnamefont {Prelovsek}},\ }\bibfield  {title} {\enquote
  {\bibinfo {title} {{Quantum spin liquid in the easy-axis Heisenberg model on
  frustrated lattices}},}\ }\href@noop {} {\bibinfo  {journal}
  {arXiv:2307.03545}\ }\BibitemShut {NoStop}%
\end{thebibliography}
%merlin.mbs apsrev4-1.bst 2010-07-25 4.21a (PWD, AO, DPC) hacked
%Control: key (0)
%Control: author (0) dotless jnrlst
%Control: editor formatted (1) identically to author
%Control: production of article title (0) allowed
%Control: page (1) range
%Control: year (0) verbatim
%Control: production of eprint (0) enabled
%

\end{document}